\documentclass[preprint,authoryear]{elsarticle}

\usepackage{amsmath, amsthm, amssymb}
\usepackage{rotfloat}
\usepackage{dsfont}

\usepackage[usenames,dvipsnames]{color}
\usepackage{multirow}
\usepackage{graphicx}
\usepackage[ruled,vlined,linesnumbered]{algorithm2e}
\usepackage{changepage}
\usepackage{hhline}
\usepackage{rotating}
\usepackage{caption}
\usepackage{hyperref}

\journal{Elsevier}

\DeclareMathOperator*{\mycup}{\cup}
\newcommand{\rmnum}[1]{\romannumeral #1}

\begin{document}

\newcommand{\PWT}{PWT}
\newcommand{\NKPc}{$\text{PWT}^c$}
\newcommand{\NKPu}{$\text{PWT}^u$}
\newcommand{\ANKP}{$\text{LB}^{\lambda}$}
\newcommand{\ANKPu}{$\text{UB}^{\lambda}$}
\newcommand{\ProfitU}{$\text{\textit{p}}^{U \! B}\!\left(x\right)$}
\newcommand{\ProfitL}{$\text{\textit{p}}^{L \! B}\!\left(x\right)$}
\newcommand{\ProfitMIP}{$\text{\textit{p}}^{M \! I \! P}\!\left(x\right)$}
\newcommand{\ProfitBIB}{$\text{\textit{p}}^{B \! I \! B}\!\left(x\right)$}
\newcommand{\ENKP}{$\text{MIP}^{\lambda}$}
\newcommand{\PWTBIB}{$\text{BIB}^{\lambda}$}
\newcommand{\SCa}{$\text{SC}^{\rmnum{1}}$}
\newcommand{\SCb}{$\text{SC}^{\rmnum{2}}$}
\newcommand{\SCc}{$\text{SC}^{\rmnum{3}}$}
\newcommand{\ANKPone}{$\text{LB}^{100}$}
\newcommand{\ANKPtwo}{$\text{LB}^{100}_{\text{\textit{no pre-pr.}}}$}
\newcommand{\ANKPthree}{$\text{LB}^{100}_{\text{\textit{red. pre-pr.}}}$}
\newcommand{\ANKPh}{$\text{NKP}_{100}^{a}$}
\newcommand{\ANKPt}{$\text{NKP}_{1000}^{a}$}
\newcommand{\ENKPone}{$\text{MIP}^{1000}$}
\newcommand{\PWTBIBone}{$\text{BIB}^{500}$}
\newcommand{\PWTBIBtwo}{$\text{BIB}^{1000}$}
\newcommand{\PWTBIBthree}{$\text{BIB}^{1000}_{\text{\textit{no seq.}}}$}
\newcommand{\PWTBIBfour}{$\text{BIB}^{1500}$}
\newcommand{\U}{uncorr}
\newcommand{\USW}{uncorr-s-w}
\newcommand{\BSC}{b-s-corr}

\newtheorem{Th}{Theorem}
\newdefinition{Def}{Definition}
\newproof{Pf}{Proof}
\theoremstyle{plain}
\newtheorem{Prop}{Proposition}

\newcommand{\remark}[1]{\textcolor{red}{#1}}

\begin{frontmatter}

\title{The Packing While Traveling Problem}

\author[a]{S. Polyakovskiy\corref{cor}} \ead{sergey.polyakovskiy@adelaide.edu.au}
\author[a]{F. Neumann} \ead{frank.neumann@adelaide.edu.au}

\cortext[cor]{Corresponding author}

\address[a]{School of Computer Science \\ The University of Adelaide \\ Adelaide, SA 5005, Australia.}

\begin{abstract}
This paper introduces the Packing While Traveling problem as a new non-linear knapsack problem. Given are a set of cities that have a set of items of distinct profits and weights and a vehicle that may collect the items when visiting all the cities in a fixed order. Each selected item contributes its profit, but produces a transportation cost relative to its weight. The problem asks to find a subset of the items such that the total gain is maximized. We investigate constrained and unconstrained versions of the problem and show that both are $\mathcal{NP}$-hard. We propose a pre-processing scheme that decreases the size of instances making them easier for computation. We provide lower and upper bounds based on mixed-integer programming (MIP) adopting the ideas of piecewise linear approximation. Furthermore, we introduce two exact approaches: one is based on MIP employing linearization technique, and another is a branch-infer-and-bound (BIB) hybrid approach that compounds the upper bound procedure with a constraint programming model strengthened with customized constraints. Our experimental results show the effectiveness of our exact and approximate solutions in terms of solution quality and computational time.
\end{abstract}


\end{frontmatter}
\section{Introduction}

Generally, the traditional statements of routing problems studied in the operations research literature base the computation of transportation costs on a linear function. However, in real practice, it might be necessary to deal with costs that have a nonlinear nature. For example, the study on the factors affecting truck fuel economy published by \cite{GOODYEAR} reveals that vehicle miles per gallon decreases as gross combination weight increases assuming speed is maintained constant. In other words, a heavily loaded truck will use much more fuel than a lightly loaded one, and this relation is not linear. 

In recent years, the research on dependence of fuel consumption on different factors, like a travel velocity, a load's weight, and vehicle's technical specifications, in various Vehicle Routing Problems (VRP) has gained attention from the operations research community. Mainly, this interest is motivated by a wish to be more accurate with the evaluation of transportation costs, and therefore to stay closer to reality. Indeed, an advanced precision would immediately benefit to transportation efficiency measured by the classic petroleum-based costs and the novel greenhouse gas emission costs. Furthermore, the proper estimation of costs and its computational simplicity should evolve optimization approaches and enhance their performance. In VRP in general, and in the Green Vehicle Routing Problems (GVRP) that consider energy consumption in particular, given are a depot and a set of customers that are to be served by a set of vehicles collecting (or delivering) required items. While the set of items is fixed, the goal is to find a route for each vehicle such that the total size of assigned items does not exceed the vehicle's capacity and the total transportation cost over all vehicles is minimized. We refer to the book of \cite{toth2002} and the recent surveys of \cite{Laporte2009} and \cite{Lin14} for an extended overview on VRP and GVRP. 

Oppositely to VRP and GVRP, we consider the situation of a single vehicle whose route is given, but items can be either collected or skipped. This situation gives rise to a problem that we designate as Packing While Traveling ({\PWT}). In the {\PWT}, the items are distributed among the cities. The vehicle visits all the cities in a specific order and collects the items of its choice. Each item has a profit and a weight, and the vehicle may collect any unless the total weight of chosen items exceeds the vehicle's capacity. The vehicle travels between two cities with a velocity that depends on the weight of the items collected in all the previously attended cities. Being selected, an item contributes its profit to the overall reward. However, its weight slows down the vehicle. This leads to a transportation cost depended on a traveling time, and therefore has a negative impact on the reward. The problem asks to find a packing plan that maximizes the difference between the total profit of the selected items and their transportation cost. The {\PWT} arises as a baseline problem in some practical applications. For example, a supplier having a single truck has to decide on goods to purchase going through a fixed route in order to maximize profitability of future sales. Specifically, there might exist only a single major route that a vehicle has to follow while any deviations from it in order to visit particular cities on the way may be negligible with respect to the length of the route. The importance of items may be variable and affected by a specific demand, therefore the profits of items can be altered accordingly from trip to trip. Obviously, {\PWT} is a part of a larger picture as a potential subproblem in various VRP with non-linear costs. Indeed, the objective function of the {\PWT} studied here is constructed similarly to one that may be based on the dependence between a fuel cost to drive a distance unit and a gross combination weight as provided in \cite{GOODYEAR}.

The {\PWT} originates from the Traveling Thief Problem (TTP) introduced by \cite{BonyadiMB13}. The TTP combines the classical Traveling Salesperson Problem (TSP) with the 0-1 Knapsack Problem (KP) and allows permutation of the order of the cities. The {\PWT} uses the same cost function as the TTP, and the only difference is the assumption of a fixed route. In this sense, any approach to the {\PWT} can also be applied to the TTP as a subroutine to solve its packing component. The TTP has been introduced and studied mainly by the evolutionary computation community during the last few years. A benchmark set based on the TSP and KP instances has been presented in \cite{Polyakovskiy14}. Approaches to handle the TTP include various meta-heuristics such as evolutionary algorithms, randomized local search and co-evolutionary approaches. \cite{Mei2015} solve the problem approximately with the two-stage memetic algorithm, which consists of a tour improvement stage and an item picking stage. For the former stage, the local search operators have been adopted different to those that are traditional for the TSP. The second stage is solved either by a constructive heuristic or by means of genetic programming. \cite{Mei2016} propose two meta-heuristics for TTP: one is the cooperative co-evolution approach that solves the sub-problems separately and transfers the information between them in each generation, and another is the memetic algorithm that solves TTP as a whole. \cite{Faulkner2015} provide multiple constructive heuristics where solutions are obtained by finding a near-optimal TSP tour by applying the Chained Lin-Kernighan heuristic (\cite{chainedLK03applegate}) to the underlying TSP part first, and then selecting a subset of items heuristically. The results produced by the heuristics have been compared then to the same approach where items are selected now by the approximate MIP-based approach of \cite{Polyakovskiy15}. They conclude that, when giving the same time limit, constructive heuristics perform generally better since are able to check more TSP tours as the packing part can be solved much faster than the MIP-based approach does. Indeed, existing approaches solve the TTP by fixing one of the components, usually the TSP, and then tackling the KP. \cite{Lourenco2016} follow in a different direction and propose an evolutionary algorithm that addresses both sub-problems at the same time. Their experimental results show that solving the TTP as a whole creates conditions for discovering solutions with enhanced quality, and that fixing one of the components might compromise the overall results. Recently, the formulation of the TTP, which allows to skip cities and to visit one city by multiple thieves, has been investigated by \cite{Chand2016}. To our best knowledge, no exact approach has been proposed for the TTP so far.

\cite{Vansteenwegen20111} give a review on the so-called orienteering problem that is somehow related to TTP. There a set of vertices is given, each with a score, and the goal is to determine a path, limited in length, that visits some vertices and maximizes the sum of the collected scores. \cite{Feillet} present a classification of traveling salesman problems with profits (TSPs with profits) and survey the existing literature on this field. TSPs with profits are a generalization of the TSP where it is not necessary to visit all vertices. A profit is associated with each vertex. The overall goal is the simultaneous optimization of the collected profit and the travel costs. In this sense, TTP has some relation to the Prize Collecting TSP~(\cite{Balas89}) where a decision is to be made on whether to visit a given city. In the Prize Collecting TSP, a city-dependent reward is obtained when a city is visited and a city-dependent penalty has to be paid for each non-visited city. In contrast to this, TTP requires that each given city is visited. Furthermore, each city has a set of available items with weights and profits and a decision has to be made on which items to pick. TTP also relates to the traveling salesman subtour problem studied by~\cite{Westerlund20062212}, where given is an undirected graph with edge costs and both revenues and weights on the vertices, and the goal is to find a subtour that includes a depot vertex, satisfies a knapsack constraint on the vertex weights, and that minimizes edge costs minus vertex revenues along the subtour. \cite{Beham2015} discuss optimization strategies for integrated knapsack and traveling salesman problems and study the Lagrangian decomposition to the knapsack constrained profitable tour problem.

In substance, the {\PWT} considers a trade-off between the profits of collected items and the transportation cost affected by their total weight. It represents a class of nonlinear knapsack problems. Knapsack problems belong to the core combinatorial optimization problems and have been frequently studied in the literature from the theoretical as well as experimental perspective~(\cite{GareyJ79,Martello90,KelPfePis04}). While the classical knapsack problem asks for maximization of a linear pseudo-Boolean function under a single linear constraint, different generalizations and variations have been investigated such as the multiple knapsack problem (\cite{ChekuriK05}) and multi-objective knapsack problems (\cite{ErlebachKP01}). Furthermore, knapsack problems with nonlinear objective functions have been studied in the literature from different perspectives (\cite{BretthauerS02}). \cite{Hochbaum95} considers the problem of maximizing a separable concave objective function subject to a packing constraint and provided an FPTAS. An exact approach for a nonlinear knapsack problem with a nonlinear term penalizing the excessive use of the knapsack capacity has been given by ~\cite{Elhedhli05}.

Recently, \cite{Junhua2016} have investigated the role of the rent rate in the {\PWT}, which is an important parameter in combining the total profit of selected items and the associated transportation cost. Specifically, the rent rate is a constant defining how much one needs to pay per unit time of traveling by the vehicle. The product of the rent rate and the total traveling time constitutes the transportation cost. In the TTP, when the value of the rent rate is small, searching for an efficient solution of the knapsack component becomes prioritized. From another hand, when the value of the rent rate is large, the TSP part of the TTP starts to play a dominating role. In this paper, the rent rate is formally introduced in Section~\ref{sec:prob} along with the {\PWT}'s statement. The theoretical and experimental investigations of \cite{Junhua2016} show how the values of the rent rate influence the difficulty of a given problem instance through the items that can be excluded by the pre-processing scheme presented in this research. Furthermore, their investigations show how to create instances that are hard to be solved by simple evolutionary algorithms. The preliminary version of our study on the {\PWT} has appeared in \cite{Polyakovskiy15}. Here, we significantly improve our earlier results. We introduce an upper bound technique that along with the enhanced pre-processing scheme allows to solve a lager range of instances to optimality and to dramatically decrease running times. Furthermore, we introduce a hybrid approach that combines constraint programming with the upper bound procedure. It is superior on many test instances and produces optimal results in a very short time. 
 
The rest of the paper is organized as follows. We give the formal statement of the {\PWT} in Section~\ref{sec:prob} and discuss its complexity in Section~\ref{sec:COMPL}. Section~\ref{sec:constr} addresses sequencing constraints that are repeatedly used later on in our approaches. In Section~\ref{sec:RS}, we provide a pre-processing scheme which allows to identify unprofitable and compulsory items, and therefore decrease the size of the {\PWT}'s instances. Section~\ref{sec:lb} explains lower and upper bound techniques. In Sections~\ref{sec:ES} and \ref{sec:BIB}, we introduce our two exact approaches: one that is based on MIP, and a hybrid one that adopts a branch-infer-and-bound paradigm. Finally, we report on the results of our experimental investigations in Section~\ref{sec:CE} and finish with some conclusions. 

\section{Problem Statement} \label{sec:prob}

The Packing While Traveling problem can be formally stated as follows. Given is a route $N=\left(1, 2,\ldots, n+1 \right)$ as a sequence of $n+1$ unique cities and a set $M$ of $m$ items distributed among first $n$ cities. Distance $d_i>0$ between two consecutive cities $(i, i+1)$ is known, for any $1\leq i \leq n$. Every city $i$ contains a set of distinct items $M_i=\{e_{i1},\ldots,e_{im_i}\}$, \mbox{$M = \mycup_{i=1}^n M_i$}. Each item $e_{ik} \in M$ has a positive integer profit $p_{ik}$ and a weight $w_{ik}$, $1 \leq k \leq m_i$. There is a single vehicle that visits all the cities in the order of a route $N$. The vehicle may collect any item in any city unless the total weight of selected items exceeds its capacity $W$. Collecting an item $e_{ik}$ leads to a profit contribution $p_{ik}$, but increases the transportation cost as the weight $w_{ik}$ slows down the vehicle. The vehicle travels along $(i, i+1)$ with velocity $v_i \in [\upsilon_{\min}, \upsilon_{\max}]$ which depends on the weight of the items collected in the first $i$ cities. When the vehicle is empty, it runs with its maximal velocity $\upsilon_{max}$. And vice verse, it runs with minimal velocity $\upsilon_{min}>0$ when is completely full. The objective is to find a subset of $M$ such that the difference between the profit of the selected items and the transportation cost is maximized.

To state the problem precisely, we give a nonlinear binary integer program formulation. Let a binary decision vector $x \in \left\{0,1\right\}^m$ represent a solution of the problem such that $x_{ik}=1$ iff $e_{ik}$ is selected. Then the travel time $t_i = \frac{d_i}{v_i}$ along $(i, i+1)$ is the ratio of the distance $d_i$ and the current velocity 
$$\upsilon_i=\upsilon_{max}-\nu \sum_{j=1}^i \sum_{k=1}^{m_j} w_{jk} x_{jk}$$ 
which is determined by the weight of the items collected in cities $1, \ldots, i$. The value $\nu = \frac{\upsilon_{max}-\upsilon_{min}}{W}$ is constant and defined by the input parameters. The velocity  depends on the weight of the chosen items linearly. The overall transportation cost is given by the sum of the transportation costs along all the edges $(i, i+1)$, $1 \leq i \leq n$, multiplied by a given rent rate $R>0$.
In summary, the problem is given by the following nonlinear binary program ({\NKPc}):

{\footnotesize
\begin{flalign}
\mbox{max} &  \displaystyle\sum_{i=1}^n \left(\displaystyle\sum_{k=1}^ {m_i} p_{ik} x_{ik}
- \frac{Rd_i}{\upsilon_{max}-\nu \displaystyle\sum_{j=1}^i \displaystyle\sum_{k=1}^{m_j} w_{jk} x_{jk}}\right) \label{eq:1}
\\ \nonumber
\mbox{s.t.} & \displaystyle\sum_{i=1}^n \displaystyle\sum_{k=1}^{m_i} w_{ik} x_{ik} \leq W 
\\ \nonumber
& x_{ik} \in \left\{0,1\right\}, \; e_{ik} \in M
\end{flalign}}

\noindent Here, (\ref{eq:1}) is a non-monotone submodular function.

We also consider the unconstrained version {\NKPu} of {\NKPc} where $W \geq \sum_{e_{ik} \in M} w_{ik}$ such that every selection of items yields a feasible solution. Given a real value $B$, the decision variant of {\NKPc} and {\NKPu} has to answer the question whether the value of (\ref{eq:1}) is at least $B$.

\section{Complexity of the Problem} \label{sec:COMPL}

In this section, we investigate the complexity of {\NKPc} and {\NKPu}. {\NKPc} is $\mathcal{NP}$-hard as it is a generalization of the classical $\mathcal{NP}$-hard 0-1 knapsack problem (\cite{Martello90}). In fact, assigning zero either to the rate $R$ or to every distance value $d_i$ in {\NKPc}, we obtain the KP. We demonstrate that in contrast to the KP the unconstrained version {\NKPu} of the problem remains $\mathcal{NP}$-hard. We show this by reducing the $\mathcal{NP}$-complete \textit{subset sum problem} (SSP) to the decision variant of {\NKPu} which asks whether there is a solution with objective value at least $B$. The input for SSP is given by $m$ positive integers $S=\left\{s_1, \ldots, s_m\right\}$ and a positive integer $Q$. The question is whether there exists a vector $x \in \left\{0,1\right\}^m$ such that $\sum_{k=1}^m {s_kx_k} = Q$.

\begin{Th}
{\NKPu} is $\mathcal{NP}$-hard.
\end{Th}

\begin{Pf} 
We start with encoding the instance of SSP given by the set of integers $S$ and the integer $Q$ as the instance $I$ of {\NKPu} having two cities. The first city contains all the $m$ items while the second city is a destination point free of items. We set the distance between two cities as $d_1=1$, the capacity of the vehicle as $W = \sum_{k=1}^m s_k$, and set $p_{1k}=w_{1k}=s_k$, $1 \leq k \leq m$. Subsequently, we set $\upsilon_{max}=2$ and $\upsilon_{min}=1$ which implies $\nu = 1/W$ and define $R^*=W\left(2-Q/W\right)^2$.

Consider the nonlinear function $f_{R^*} \colon \left[0,W\right] \rightarrow \mathbb{R}$ defined as
{\footnotesize
\begin{align}
\nonumber f_{R^*}\left(w\right)=w-\frac{R^*}{2- w/W}.
\end{align}}
\noindent $f_{R^*}$ defined on the interval $\left[0,W\right]$ is a continuous concave function that reaches its unique maximum in the point $w^*=W \cdot (2-\sqrt{ R^*/W}) = Q$ , i.e. \mbox{$f_{R^*}\left(w \right)<f_{R^*}\left(w^*\right)$} for $w \in [0,W]$ and $w \not = w^*$. Then $f_{R^*}(Q)$ is the maximum value for $f_{R^*}$ when being restricted to integer input, too. Therefore, we set $B=f_{R^*}(Q)$ and the objective function for {\NKPu} is given by
{\footnotesize
\nonumber \begin{align}
g_{R^*}\left(x\right)=\displaystyle\sum_{k=1}^m {p_{1k}x_k}-\frac{R^*}{2- \frac{1}{W} \displaystyle\sum_{k=1}^m {w_{1k}x_k}}.
\end{align}}
There exists an $x \in \{0,1\}^m$ such that $g_{R^*}(x) \geq B= f_{R^*}(Q) = 2\left(Q-W\right)$ iff $\sum_{k=1}^m s_kx_k=\sum_{k=1}^m w_{1k}x_k=\sum_{k=1}^m p_{1k}x_k=Q$. Therefore, the instance of SSP has answer YES iff the optimal solution of the {\NKPu} instance $I$ has objective value at least $B=f_{R^*}\left(Q\right)$. Obviously, the reduction can be carried out in polynomial time which completes the proof.
\qed
\end{Pf}

\section{Sequencing Constraints} \label{sec:constr}

In this section, we derive a set of constraints that speed up the reasoning to be done within our algorithms. Specifically, the constraints establish priority among the items positioned in the same or different cities of the route. 

The first portion of the constraints results from the fact that item $e_{il}$ in city $i$ should not be selected prior to another item $e_{ik}$, $1 \leq l,k \leq m_i$ and $l \neq k$, positioned in the same city when the condition $\left(p_{il}<p_{ik}\right) \wedge \left(w_{il}\geq w_{ik}\right)$ holds. This constraint ({\SCa}) has the form of
{\footnotesize
\begin{flalign}
\nonumber&x_{il}\leq x_{ik}, \; l \neq k, \; e_{il},e_{ik} \in M_i \;:\; \left(p_{il}<p_{ik}\right) \wedge \left(w_{il} \geq w_{ik}\right).
\end{flalign}
}

\noindent Let $\Delta_l^{ji}$ denote a lower bound on the cost of transportation of item $e_{jl}$ from city $j$ to succeeding city $i$ computed as
{\footnotesize
\begin{equation}
\nonumber\Delta_l^{ji}=R \sum_{a=j}^{i-1} d_a\left(\frac{1}{\upsilon_{max}-\nu \left(w_{jl}+\displaystyle\sum_{b=1}^a w_b^c\right)}-\frac{1}{\upsilon_{max}-\nu \displaystyle\sum_{b=1}^a w_b^c}\right),
\end{equation}}
\noindent where $w_b^c$ is the total weight of the compulsory items collected in city $b$. Compulsory items must be a part of an optimal solution (see Section \ref{sec:RS} for details). Specifically, $\Delta_l^{ji}$ is based on the difference between traveling with all the compulsory items and item $e_{jl}$ and traveling with all the compulsory items only. In this case, no other items are picked up and the vehicle runs with the maximal feasible velocity that it may achieve on each of the edges. Then another set of constraints uses the fact that item $e_{jl}$ in city $j$ should not be selected prior to item $e_{ik}$ in city $i$ such that the condition $\left(p_{jl}-\Delta_l^{ji}<p_{ik}\right) \wedge \left(w_{jl}\geq w_{ik}\right)$ holds. This constraint ({\SCb}) takes the form of
{\footnotesize
\begin{flalign}
\nonumber&x_{jl}\leq x_{ik}, \; j<i, \; e_{jl}\in M_j, \; e_{ik} \in M_i \;: \; \left(p_{jl}-\Delta_l^{ji}<p_{ik}\right) \wedge \left(w_{jl}\geq w_{ik}\right).
\end{flalign}
}

Similarly, let $\overline{\Delta}_l^{ji}$ denote an upper bound on the cost of transportation of item $e_{jl}$ from city $j$ to succeeding city $i$ computed as
{\footnotesize
\begin{equation}
\nonumber\overline{\Delta}_l^{ji}=R\sum_{a=j}^{i-1} d_a \left(\frac{1}{\upsilon_{max}-\nu\cdot min\left(\displaystyle\sum_{b=1}^a w_b^{max},W\right)}-\frac{1}{\upsilon_{max}-\nu \left( min\left(\displaystyle\sum_{b=1}^a w_b^{max},W\right)-w_{jl}\right)}\right),
\end{equation}
}
\noindent where $w_b^{max}$ is the total weight of all the items existing in city $b$. Specifically, $\overline{\Delta}_l^{ji}$ represents the difference between traveling having the vehicle loaded as much as possible and doing so but leaving a free space just for item $e_{jl}$. In this case, as many items as possible are picked up and the vehicle travels with the least feasible velocity that it may achieve on each of the edges. Then one more set of constraints arises from the fact that item $e_{ik}$ in city $i$ should not be selected prior to item $e_{jl}$ in city $j$ such that $\left(p_{jl}-\overline{\Delta}_l^{ji}>p_{ik}\right) \wedge \left(w_{jl} \leq w_{ik}\right)$ holds. This constraint ({\SCc}) has the following form:
{\footnotesize
\begin{flalign}
\nonumber&x_{jl}\geq x_{ik}, \; j<i, \; e_{jl}\in M_j, \; e_{ik} \in M_i \;: \; \left(p_{jl}-\overline{\Delta}_l^{ji}>p_{ik}\right) \wedge \left(w_{jl}\leq w_{ik}\right).
\end{flalign}
}

\section{Pre-processing} \label{sec:RS}
In this section, we introduce a pre-processing scheme to identify items of a given instance $I$ that can be either directly included or discarded. Excluding such items from solution process can significantly speed up algorithms. We distinguish between two kinds of items that are identified in the pre-processing: \emph{compulsory} and \emph{unprofitable} items. We call an item \emph{compulsory} if its inclusion in any feasible solution increases the objective function value, and call an item \emph{unprofitable} if it does not do that. Therefore, an optimal solution must contain all compulsory items while all unprofitable items must be discarded. In order to identify \emph{compulsory} and \emph{unprofitable} items, we consider the total transportation cost that a set of items produces.

\begin{Def}[Total Transportation Cost]
Let $O \subseteq M$ be a subset of items. We define the total transportation cost along route $N$ when the items of $O$ are selected as
{\footnotesize
\begin{align} 
\nonumber t_O= R \cdot \sum_{i=1}^n  \frac{d_i}{\upsilon_{max}-\nu \sum_{j=1}^i \sum_{e_{jk} \in O_j} w_{jk}},
\end{align}}
\noindent where $O_j = M_j \cap O$, $1 \leq j \leq n$, is the subset of $O$ selected in city $j$.
\end{Def}

Based on the given instance $I$, we can identify unprofitable items for {\NKPc} according to the following proposition.

\begin{Prop}[\textbf{Unprofitable Item, {\NKPc} Case}] \label{prop1}
Let $I$ be an arbitrary instance of {\NKPc}. If $p_{ik} \leq R\left(t_{\{e_{ik}\}} - t_\emptyset \right)$, then $e_{ik}$ is an unprofitable item.
\end{Prop}

\begin{Pf} 
We assume that $p_{ik}\leq R\left(t_{\left\{e_{ik}\right\}}-t_\emptyset\right)$ holds. Let $M^* \subseteq M \setminus \left\{e_{ik} \right\}$ denote an arbitrary subset of items excluding $e_{ik} $ such that \mbox{$w_{ik}+\sum_{e_{jl}\in M^*} w_{jl} \leq W$} holds. We consider $t_{M^*\cup \left\{ e_{ik} \right\} }$ and $t_{M^*}$. Since the velocity depends linearly on the weight of collected items and the travel time $t_i=d_i/v_i$ along $(i,i+1)$ depends inversely proportional on the velocity $v_i$, the inequality $\left(t_{\left\{e_{ik}\right\}}-t_\emptyset\right) \leq \left(t_{M^*\cup \left\{ e_{ik} \right\} }-t_{M^*}\right)$ holds. Therefore, $p_{ik}\leq R\left(t_{M^*\cup \left\{ e_{ik} \right\}}-t_{M^*}\right)$ holds for any $M^* \subseteq M \setminus \left\{e_{ik} \right\}$ that completes the proof.
\qed
\end{Pf}

Proposition~\ref{prop1} helps to determine whether the profit $p_{ik}$ of item $e_{ik}$ is large enough to cover the least incremental transportation cost it incurs when selected in the packing plan $x$. In this case, the least incremental transportation cost results from accepting the selection of $e_{ik}$ as only selected item in $x$ versus accepting empty $x$ as a solution. It is important to note that Proposition~\ref{prop1} can reduce {\NKPc} problem to {\NKPu} by excluding items so that the sum of the weights of all remaining items does not exceed the weight bound $W$. In this case, we can further refine the set of items by searching for those ones that must be a part of any solution of {\NKPc}. We identify compulsory items for the unconstrained case according to the following proposition.

\begin{Prop}[\textbf{Compulsory Item, {\NKPu} Case}] \label{prop2}
Let $I$ be an arbitrary instance of {\NKPu}.
If $p_{ik} > R\left(t_{M} - t_{M \setminus \left\{ e_{ik} \right\} } \right)$, then $e_{ik}$ is a compulsory item.
\end{Prop}

\begin{Pf}
We work under the assumption that $p_{ik} > R\left(t_{M}-t_{M \setminus \left\{ e_{ik} \right\} }\right)$ holds. In the case of {\NKPu}, all the existing items can fit into the vehicle at once and all subsets $O \subseteq M$ are feasible. Let $M^* \subseteq M \setminus \left\{ e_{ik} \right\}$ be an arbitrary subset of items excluding $e_{ik}$, and consider $t_{M \setminus M^*}$ and $t_{M \setminus M^* \setminus \left\{ e_{ik}\right\} }$, respectively. Since the velocity depends linearly on the weight of collected items and the travel time $t_i=d_i/v_i$ along $(i,i+1)$ depends inversely proportional on the velocity $v_i$, we have $\left(t_{M}-t_{M \setminus \left\{ e_{ik} \right\} }\right) \geq \left(t_{M \setminus M^*}-t_{M \setminus M^* \setminus \left\{ e_{ik}\right\} }\right)$. This implies that $p_{ik}>R\left(t_{M \setminus M^*}-t_{M \setminus M^* \setminus \left\{ e_{ik}\right\} }\right)$ holds for any subset $M \setminus M^*$ of items which completes the proof.
\qed
\end{Pf}

For the unconstrained variant {\NKPu}, Proposition~\ref{prop2} is valid to determine whether item $e_{ik}$ is able to cover by its $p_{ik}$ the largest possible incremental transportation cost it may generate when has been selected in $x$. Here, the largest possible incremental transportation cost is computed with respect to the worst case scenario when all the possible items are selected along with $e_{ik}$, and therefore the vehicle has the maximal possible load and the least velocity, versus accepting all the items but $e_{ik}$. In the unconstrained case, having compulsory items included according to Proposition~\ref{prop2}, we may identify some more unprofitable items. Indeed, compulsory items contribute to the collected weight and therefore limit the potential positive contribution of other items. As a result, some of the items may become unprofitable after a number of compulsory items has been detected. We find unprofitable items for {\NKPu} with respect to the following proposition.

\begin{Prop}[\textbf{Unprofitable Item, {\NKPu} Case}] \label{prop3}
Let $I$ be an arbitrary instance of {\NKPu} and $M^c$ be the set of all compulsory items. If $p_{ik} \leq R\left(t_{M^c \cup \left\{e_{ik}\right\} } - t_{M^c} \right)$, then $e_{ik}$ is an unprofitable item.
\end{Prop}

\begin{Pf}
We assume that $p_{ik} \leq R\left(t_{M^c \cup \left\{e_{ik}\right\} }-t_{M^c} \right)$ holds. Let $M^* \subseteq M \setminus \left\{ M^c \cup \left\{e_{ik}\right\} \right\}$ be an arbitrary subset of $M$ that does not include any item of  $M^c \cup \left\{e_{ik} \right\}$ and consider $t_{M^c \cup M^*}$ and $t_{M^c \cup M^* \cup \left\{ e_{ik} \right\} }$.
Since the velocity depends linearly on the weight of collected items and the travel time $t_i=d_i/v_i$ along $(i,i+1)$ depends inversely proportional on the velocity $v_i$, we have
$\left( t_{M^c \cup \left\{e_{ik}\right\} }-t_{M^c} \right) \leq \left(t_{M^c \cup M^* \cup \left\{ e_{ik} \right\} }-t_{M^c \cup M^*} \right)$. Hence, we have $p_{ik} \leq R\left(t_{M^c \cup M^* \cup \left\{e_{ik} \right\}}-t_{M^c \cup M^*} \right)$ for any $M^*\subseteq M \setminus \left\{ M^c \cup \left\{e_{ik}\right\} \right\}$ which completes the proof.
\qed
\end{Pf}

Proposition~\ref{prop3} determines for {\NKPu} whether the profit $p_{ik}$ of item $e_{ik}$ is large enough to cover the least incremental transportation cost resulted from its selection along with all known compulsory items. Specifically, in Proposition~\ref{prop3} the list incremental transportation cost follows from accepting the selection of $e_{ik}$ along with the set of compulsory items $M^c$ in $x$ versus accepting just the selection of $M^c$ as a solution. One can see that Proposition~\ref{prop3} is a special case of Proposition~\ref{prop1} with only the difference that it has $t_\emptyset$ replaced by $t_{M^c}$.

It takes only a linear time to check any instance of {\NKPc} for unprofitable items with respect to Proposition~\ref{prop1}. In fact, each item $e_{ik}$ can be checked in a constant time if the total length of the path from city $i$ to city $n+1$ is known. When dealing with {\NKPu}, Propositions~\ref{prop2} and \ref{prop3} can be applied iteratively to the remaining set of items until no compulsory or unprofitable item is found. The running time of all the rounds of the search is bounded by $\mathcal{O}\left(nm^2\right)$. Our preliminary investigation has shown that it is rather time-consuming to solve large and even moderate-sized unconstrained instances due to the time spent on computing the incremental transportation cost for each of the items separately as the Propositions~\ref{prop2} and \ref{prop3} advise. Indeed, we cannot perform the pre-processing step reasonably fast with respect to the time limits we apply in Section \ref{sec:CE}. Obviously, slow pre-processing can easily stultify all benefits of its use. To manage this, we use the reasoning similar to one that the sequencing constraints adopt in Section \ref{sec:constr}. Specifically, we deduce whether item $e_{ik}$ is compulsory or unprofitable from the answer concerning item $e_{jl}$ for which it has been already obtained. Algorithm \ref{alg:PP} sketches the pseudocode of the enhanced algorithm, which runs in $\mathcal{O}\left(m^3\right)$, but operates up to two orders of magnitude faster in practice and allows to handle the largest instances of the test suite (see Section \ref{sec:CE} for details). 

The pre-processing algorithm works as follows. The loop (\ref{l5}-\ref{l24}) searches for compulsory items and the loop (\ref{l29}-\ref{l52}) searches for unprofitable ones. Once either no compulsory or no unprofitable item has been found within the corresponding loop, the algorithm terminates (cf. lines \ref{l25} and \ref{l53}). We use two Boolean variables $\mu_{ik}^u$ and $\mu_{ik}^c$ that take value $true$ to mark item $e_{ik}$ as unprofitable and compulsory, respectively. Both values are initialized as $\mu_{ik}^u=\mu_{ik}^c=false$. Subsequently, variable $\overline{w}_i^{max}=\sum_{b=1}^i w_b^{max}$ computes the maximal possible weight of the items that can be collected in city $i$ and in all the preceding cities. Similarly, variable $\overline{w}_i^c=\sum_{b=1}^i w_b^c$ computes the total weight of compulsory items existing in city $i$ and in all the preceding cities. We use $\overline{w}_i^{max}$ and $\overline{w}_i^c$ to calculate, respectively, the largest possible incremental cost $c_{max}$ and the minimal possible incremental cost $c_{min}$ for each of the items in the loops of our algorithm (cf. lines \ref{l7}-\ref{l24} and lines \ref{l31}-\ref{l52}). Furthermore, to make reasoning on the properties of item $e_{ik}$ with respect to the known properties of item $e_{jl}$, we introduce the dummy profit $p'_{jl}$ of $e_{jl}$. Specifically, when item $e_{jl}$ has been shown to be either compulsory or not, or either unprofitable or not, $p'_{jl}$ defines how large or small profit $p_{ik}$ must be with respect to computed $c_{max}$ or $c_{min}$ to let $e_{ik}$ have the same property as $e_{jl}$ (cf. lines \ref{l11}, \ref{l14}, \ref{l35}, and \ref{l38}). For example, when item $e_{ik}$ is proved to be compulsory independently of any another item (cf. line \ref{l24}), its $p'_{ik}$ is set to $c_{max}$ (cf. line \ref{l23}). This means that $e_{ik}$ would be compulsory even if its profit was less than $p_{ik}$, but mainly greater than $c_{max}$. Therefore, to become a compulsory item as $p_{ik}$ is, another item, say $e_{i'k'}$ in city $i': i'\leq i$, should have a weight that is smaller or equal to $w_{ik}$ and its profit $p_{i'k'}$ minus the corresponding largest possible incremental cost must be strictly larger than $p'_{ik}$ (cf. line \ref{l14}). Similarly, when $e_{ik}$ is proved to be not a compulsory item independently of any other item, its $p'_{ik}$ is also set to $c_{max}$ (cf. line \ref{l23}). This is because $e_{ik}$ would not be compulsory even if its profit was larger than $p_{ik}$, but mainly less or equal to $c_{max}$. Therefore, to stay as not a compulsory item as $p_{ik}$ is, another item, say $e_{i'k'}$ in city $i': i'\leq i$, should have a weight that is greater or equal to $w_{ik}$ and its profit $p_{i'k'}$ minus the corresponding least possible incremental cost must be at most $p'_{ik}$ (cf. line \ref{l11}). The same reasoning is to be done for the case when $e_{ik}$ is proved as a compulsory item according to already known compulsory item $e_{jl}$. Here, $p'_{ik}$ is set to $p'_{jl}+c_{max}$ where $c_{max}$ plays a role of the largest cost of transportation of $e_{ik}$ from city $i$ to succeeding city $j$ (cf. line \ref{l16}). In such a way, when considering other items in the future iterations, they are compared to item $e_{jl}$ through the current item $e_{ik}$ since $e_{ik}$ implicitly points to $e_{jl}$ using the assigned dummy profit $p'_{ik}$. The same reasoning is valid for proving $e_{ik}$ to be not a compulsory item according to already known non-compulsory item $e_{jl}$ where $p'_{ik}$ is set to $p'_{jl}+c_{min}$ (cf. line \ref{l12}). In a similar way, we proceed with deduction of unprofitable items in the loop (\ref{l29}-\ref{l52}). Utilizing the dummy profits of items significantly strengthen deductions by relating the items to one of the items' group rapidly. Before applying our approaches given in Section~\ref{sec:lb}, Section~\ref{sec:ES}, and \ref{sec:BIB}, we remove all unprofitable and compulsory items from the set $M$ using these pre-processing steps.

\SetAlFnt{\tiny}
\begin{algorithm}[!htb]\caption{The Pre-processing Algorithm}\label{alg:PP}
initialize the indicator variables $\mu_{ik}^u=\mu_{ik}^c=false$ for each item $e_{ik} \in M$\;
\While{true}{
set $flag \leftarrow false$\;
\For{each city $i$ from 1 to $n$}{calculate $\overline{w}_i^{max}$\;}
\For{each city $i$ from $n$ to 1}{\label{l5}
\For{each item $e_{ik} \in M_i : \neg\left(\mu_{ik}^u \vee \mu_{ik}^c\right)$}{
$c_{max} \leftarrow 0$; $c_{min} \leftarrow 0$\;\label{l7}
initialize $flag' \leftarrow false$\;
\For{each city $j$ from $i$ to $n$}{
\For{each item $e_{jl} \in M_j: \neg \mu_{jl}^u \wedge \left(\left(i \neq j\right) \vee \left(k > l\right)\right)$}{
\If{$\left(\neg \mu_{jl}^c\right) \wedge \left(w_{ik} \geq w_{jl}\right) \wedge \left(p_{ik} - c_{min} \leq p'_{jl}\right)$}{ \label{l11}
$p'_{ik} \leftarrow p'_{jl}+c_{min}$\; \label{l12}
$flag' \leftarrow true$; break\;
}
\If{$\mu_{jl}^c \wedge \left(w_{ik} \leq w_{jl}\right) \wedge \left(p_{ik} - c_{max} > p'_{jl}\right)$}{ \label{l14}
$\mu_{ik}^c \leftarrow true$\;
$p'_{ik} \leftarrow p'_{jl}+c_{max}$\;\label{l16}
$flag \leftarrow true$\;
$flag' \leftarrow true$; break\;
}
}
\lIf{$flag'$}{break}
$c_{min} \leftarrow c_{min}+Rd_j\left(\frac{1}{\upsilon_{max}-\nu\left(\overline{w}_j^c+w_{ik}\right)}-\frac{1}{\upsilon_{max}-\nu\overline{w}_j^c}\right)$\;
$c_{max} \leftarrow c_{max}+Rd_j\left(\frac{1}{\upsilon_{max}-\nu\overline{w}_j^{max}}-\frac{1}{\upsilon_{max}-\nu\left(\overline{w}_j^{max}-w_{ik}\right)}\right)$\;
}
\lIf{$flag'$}{break}
$p'_{ik} \leftarrow c_{max}$\;\label{l23}
\lIf{$c_{max}<p_{ik}$}{ $\mu_{ik}^c \leftarrow true$; $flag \leftarrow true$} \label{l24}
}
}
\lIf{$\neg flag$}{break}\label{l25}
set $flag \leftarrow false$\;
\For{each city $i$ from 1 to $n$}{calculate $\overline{w}_i^c$\;}
\For{each city $i$ from $n$ to 1}{\label{l29}
\For{each item $e_{ik} \in M_i : \neg \left(\mu_{ik}^u \vee \mu_{ik}^c\right)$}{
$c_{max} \leftarrow 0$; $c_{min} \leftarrow 0$\;\label{l31}
initialize $flag' \leftarrow false$\;
\For{each city $j$ from $i$ to $n$}{
\For{each item $e_{jl} \in M_j: \neg \mu_{jl}^c \wedge \left(\left(i \neq j\right) \vee \left(k > l\right)\right)$}{
\If{$\left(\neg \mu_{jl}^u\right) \wedge \left(w_{ik} \leq w_{jl}\right) \wedge \left(p_{ik} - c_{max} > p'_{jl}\right)$}{ \label{l35}
$p'_{ik} \leftarrow p'_{jl}+c_{max}$\; 
$flag' \leftarrow true$; break\;
}
\If{$\mu_{jl}^u \wedge \left(w_{ik} \geq w_{jl}\right) \wedge \left(p_{ik} - c_{min} \leq p'_{jl}\right)$}{ \label{l38}
$\mu_{ik}^u \leftarrow true$\;
$p'_{ik} \leftarrow p'_{jl}+c_{min}$\;
$flag \leftarrow true$\;
$flag' \leftarrow true$; break\;
}
}
\lIf{$flag'$}{break}
$c_{min} \leftarrow c_{min}+Rd_j\left(\frac{1}{\upsilon_{max}-\nu\left(\overline{w}_j^c+w_{ik}\right)}-\frac{1}{\upsilon_{max}-\nu\overline{w}_j^c}\right)$\;
\If{$c_{min} \geq p_{ik}$}{ 
$\mu_{ik}^u \leftarrow true$\; 
$p'_{ik} \leftarrow c_{min}$\;
$flag \leftarrow true$\;
$flag' \leftarrow true$; break\;
}
$c_{max} \leftarrow c_{max}+Rd_j\left(\frac{1}{\upsilon_{max}-\nu\overline{w}_j^{max}}-\frac{1}{\upsilon_{max}-\nu\left(\overline{w}_j^{max}-w_{ik}\right)}\right)$\;
}
\lIf{$flag'$}{break}
$p'_{ik} \leftarrow c_{min}$\;\label{l52}
}
}
\lIf{$\neg flag$}{break}\label{l53}
}
\end{algorithm}

\section{Lower and Upper Bounds} \label{sec:lb}
In practice, approximation of nonlinear terms is an efficient way to deal with them. Although an approximate solution is likely to be different from an exact one, it might be close enough and obtainable in a reasonable computational time. In this section, we propose lower and upper bound techniques based on mixed-integer programming (MIP) adopting the ideas of piecewise linear approximation.

\subsection{Lower Bound} \label{sec:llb}

\emph{\begin{figure}[htb]
\centering
\includegraphics[width=\textwidth]{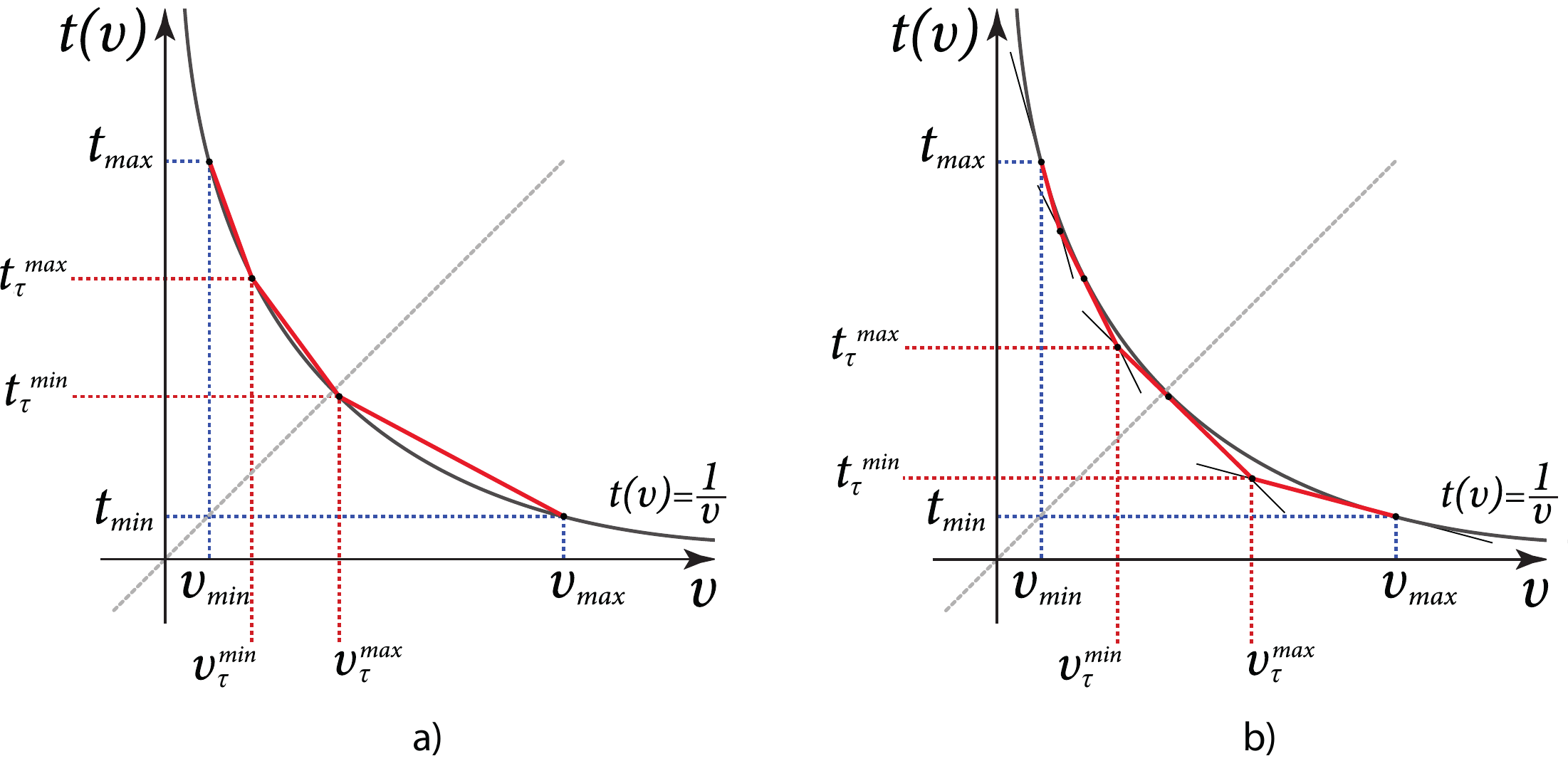}
\caption{Piecewise linear approximation of $t\left(\upsilon\right)=1/\upsilon$}
\label{fig:fig1}
\end{figure}}

Consider an arbitrary edge $\left(i,i+1\right)$ and the traveling time $t'_i \in [t_{\min}, t_{\max}]$ per distance unit along it, for any $i=1, \ldots, n$. Here, $t_{min}=1/\upsilon_{max}$ and $t_{max}=1/\upsilon_{min}$ bound $t'_i$ from below and from above, respectively. We partition the interval $\left[t_{min},t_{max}\right]$ into $\lambda$ equal-sized sub-intervals and determine thus a set $T=\left\{\tau_1,\ldots,\tau_{\lambda}\right\}$ of straight line segments to approximate the curve of the function $t\left(\upsilon\right)$ as illustrated in Figure~\ref{fig:fig1}a. Each segment $\tau \in T$ is characterized by its minimal velocity $\upsilon_\tau^{min}$ and its corresponding maximum traveling time per distance unit $t_\tau^{max}$, and by its maximum velocity $\upsilon_\tau^{max}$ and its corresponding minimum traveling time per distance unit $t_\tau^{min}$. Specifically, $\left(\upsilon_\tau^{min},t_\tau^{max}\right)$ and $\left(\upsilon_\tau^{max},t_\tau^{min}\right)$ are the endpoints of segment $\tau$ referred to as breakpoints. We approximate $t'_i$ by the linear combination of $t_\tau^{min}$ and $t_\tau^{max}$ if $\upsilon_i \in \left[\upsilon_\tau^{min},\upsilon_\tau^{max}\right]$. 

Our lower bound MIP-based model uses three types of variables in addition to the binary decision variable $x_{ik}$ for each item $e_{ik} \in M$ from Section~\ref{sec:prob}. Let $w_i$ be a real variable equal to the total weight of selected items when traveling along the $\left(i,i+1\right)$. Let $p_i$ be a real variable equal to the difference of the total profit of selected items and their total transportation cost when delivering them to city $i+1$. Let $T_i \subseteq T$, $1 \leq i \leq n$, denote a set of possible segments to which velocity $\upsilon_i$ of the vehicle may relate, i.e. $T_i = \left\{\tau \in T\; :\;
\left( \upsilon_\tau^{\min} \in \left[ \upsilon_i^{\min}, \upsilon_i^{\max} \right] \right) \vee 
\left( \upsilon_\tau^{\max} \in \left[ \upsilon_i^{\min}, \upsilon_i^{\max} \right] \right) \right\}$, where $\upsilon_i^{\max} = \upsilon_{\max}- \nu\sum_{j=1}^i w_j^c$ is the maximal possible velocity that the vehicle can move along $\left(i,i+1\right)$ when packing in all compulsory items only, and $\upsilon_i^{\min}=\upsilon_{max}- \nu \cdot \min\left(\sum_{j=1}^i w_j^{max},W\right)$ the minimum possible velocity along $(i, i+1)$ after having packed in all items available in cities $1, \ldots, i$. Actually, we have $\upsilon_i \in \left[\upsilon_i^{min}, \upsilon_i^{max}\right]$. When $\upsilon_i \in \left[ \upsilon_\tau^{\min}, \upsilon_\tau^{\max} \right]$ for $\tau \in T$, any point in between endpoints of $\tau$ is a weighted sum of them. Let $B_i$ denote a set of all breakpoints that the linear segments of $T_i$ have. Then the value of the real variable $y_{ib} \in \left[0,1\right]$ is a weight assigned to the breakpoint $b \in B_i$ associated with the pair of values $\left(\upsilon_b,t_b\right)$. When the linear combination $\sum_{b \in B_i}\upsilon_by_{ib}$ under the constraint $\sum_{b\in B_i} y_{ib} = 1$ equals $\upsilon_i$, the linear combination $\sum_{b \in B_i}t_b y_{ib}$ overestimates $t'_i$. This underestimates the resulting profit minus the total transportation cost and gives a valid lower bound for {\NKPc} (and {\NKPu}) that can be obtained by solving the following linear mixed 0-1 program ({\ANKP}):

{\footnotesize
\begin{flalign}
\mbox{max} \;  &\text{{\ProfitL}} = p_n \label{eq:a0}
\\
\mbox{s.t.} \; & p_i=p_{i-1}+p_i^c+\displaystyle\sum_{e_{ik}\in M_i} p_{ik}x_{ik}-Rd_i\displaystyle\sum_{b\in B_i} t_by_{ib}, \; i=1,\ldots,n \label{eq:a1}
\\
& w_i=w_{i-1} + w_i^c+\displaystyle\sum_{e_{ik}\in M_i} w_{ik}x_{ik}, \; i=1,\ldots,n \label{eq:a2}
\\
& \nu w_i+\displaystyle\sum_{b\in B_i} \upsilon_b y_{ib}=\upsilon_{max}, \; i=1,\ldots,n \label{eq:a3}
\\
& \displaystyle\sum_{b\in B_i} y_{ib} = 1, \; i=1,\ldots,n \label{eq:a4}
\\
& w_n\leq W \label{eq:a5}
\\  
& x_{ik} \in \left\{0,1\right\}, \; e_{ik} \in M \label{eq:a6}
\\  
& y_{ib} \in \left[0,1\right], \; i=1,\ldots,n,\;b \in B_i \label{eq:a7}
\\ 
& p_i\in \mathbb{R}, \; i=1,\ldots,n \label{eq:a8}
\\
& w_i\in \mathbb{R}_{\geq 0}, \; i=1,\ldots,n \label{eq:a9}
\\
& p_0=w_0=0 \label{eq:a10}
\end{flalign}
}

The value of $\lambda$ in {\ANKP} sets its precision. Indeed, the precision of the lower bound may be increased at the cost of a running time as this also increases the number of segments, and thus raises the number of $y$-type variables to be involved. Equation (\ref{eq:a0}) defines the objective function {\ProfitL} as $p_n$ that is the difference of the total profit of selected items delivered to city $n+1$ and their total transportation cost. Since the transportation cost is approximated in {\ANKP}, the actual objective value for {\NKPc} (and {\NKPu}) should be computed on the values of the decision variables of vector $x$. The resulting value then is also a valid lower bound. Equation (\ref{eq:a1}) computes the difference $p_i$ of the total profit of selected items and their total transportation cost when arriving at city $i+1$ by summing up the value of $p_{i-1}$ concerning $\left(i-1,i\right)$, the profit of compulsory items $p_i^c$ and the profit $\sum_{e_{ik}\in M_i} p_{ik}x_{ik}$ of items selected in city $i$, and subtracting the approximated transportation cost along $(i, i+1)$. Equation (\ref{eq:a2}) gives the weight $w_i$ of the selected items when the vehicle departs city $i$ by summing up $w_{i-1}$, the weight of compulsory items $w_i^c$ and the weight $\sum_{e_{ik}\in M_i} w_{ik}x_{ik}$ of items selected in city $i$. Remind that we determine compulsory items according to the Proposition~\ref{prop2} of Section~\ref{sec:RS} when the problem is unconstrained. Equation (\ref{eq:a3}) implicitly defines segment $\tau \in T_i$ to which the velocity of the vehicle $\upsilon_i$ belongs and sets the weights for its endpoints. Equation (\ref{eq:a4}) forces the total weight of the breakpoints of $B_i$ be exactly 1. Equation (\ref{eq:a5}) imposes the capacity constraint, and Eq. (\ref{eq:a6}) declares $x_{ik}$ as binary. Equation (\ref{eq:a7}) states $y_{ib}$ as a real variable defined in $\left[0,1\right]$. Equation (\ref{eq:a8}) declares $p_i$ as a real variable, while Eq. (\ref{eq:a9}) defines $w_i$ as a non-negative real. Finally, Equation (\ref{eq:a10}) establishes the base cases for $p_0$ and $w_0$. Obviously, one can relax the integrality imposed on the $x$-type variables that leads to a linear programming model. In fact, this generally worsens the lower bound value, but gives an advantage in running time.

\subsection{Upper Bound} \label{sec:ub}

We now describe the upper bound technique that adopts the piecewise linear approximation proposed for the lower bound. This time, our goal is to underestimate the traveling time $t'_i$ that the vehicle spends to pass a distance unit when traveling along the $\left(i,i+1\right)$, for any $i=1, \ldots, n$. We utilize the same set of breakpoints $B_i$ generated from the set of linear segments $T_i$. In each point $b \in B_i$, we draw a tangent to the curve of the function $t\left(\upsilon\right)=1/\upsilon$ as depicted in Figure~\ref{fig:fig1}b. Subsequently, a new set of points $\overline{B}_i$ is derived from the left and the rightmost points of $B_i$, and the points of intersection of each pair of neighboring tangents. This yields totally $\left|B_i\right|+1$ points that produce a new set of $\left|B_i\right|$ linear segments $\overline{T}_i$ resulted from connecting every two closest points in $\overline{B}_i$. Then a valid upper bound for {\NKPc} (and {\NKPu}) can be obtained via the model of {\ANKP} with only the difference that the set of breakpoints $\overline{B}_i$ is used instead of $B_i$. We designate this altered model as {\ANKPu} and the corresponding objective function as {\ProfitU}. Again, one can manage precision of the upper bound adjusting the value of $\lambda$. Furthermore, the integrality imposed on the $x$-type variables may be relaxed to speed up computations at the price of the upper bound's quality. 

\section{Mixed-Integer Programming-Based Approach} \label{sec:ES}

Both {\NKPc} and {\NKPu} belong to the specific class of fractional binary programming problems for which several efficient reformulation techniques exist to handle nonlinear terms. We follow the approach of \cite{Li94} and \cite{Tawarmalani02} to reformulate {\NKPc} (and {\NKPu}) as a linear mixed 0-1 program. It is applicable since the denominator of each fractional term in (\ref{eq:1}) is not equal to zero since $\upsilon_{min}>0$. We start with introduction of auxiliary real-valued variables $y_i$, $i=1, \ldots, n$, such that $y_i = 1/\left(\upsilon_{max}-\nu \sum_{j=1}^i \sum_{k=1}^{m_j} w_{jk} x_{jk} \right)$. The variables $y_i$ express the travel time per distance unit along the edge $\left(i, i+1\right)$. According to \cite{Li94}, we can reformulate {\NKPc} as a mixed 0-1 quadratic program by replacing (\ref{eq:1}) with (\ref{eq:Li1}) and adding the set of constraints (\ref{eq:Li2}) and (\ref{eq:Li3}).

{\footnotesize
\begin{flalign}
\mbox{max} & \displaystyle\sum_{i=1}^n \left( \displaystyle\sum_{k=1}^{m_i} p_{ik} x_{ik} - Rd_i y_i\right)\label{eq:Li1}
\\ 
\mbox{s.t.} \;& \upsilon_{max}y_i + \nu \displaystyle\sum_{j=1}^i \displaystyle\sum_{k=1}^{m_j} w_{jk} x_{jk} y_i = 1, \; i=1,\ldots,n \label{eq:Li2}
\\ 
& y_i \in \mathbb{R}_+, \; i=1,\ldots,n \label{eq:Li3}
\end{flalign}
}

According to \cite{Tawarmalani02}, if $z=xy$ is a polynomial mixed 0-1 term where $x$ is binary and $y$ is a real-valued variable, then it can be linearized via the set of linear inequalities: (i) $z\leq Ux$;  (ii) $z\geq Lx$;  (iii) $z\leq y + L\left(x-1\right)$; (iiii) $z\geq y + U\left(x-1\right)$. Here, $U$ and $L$ are the upper and lower bounds on $y$, i.e. $L\leq y\leq U$. We can linearize the $x_{jk} y_i$ term in (\ref{eq:Li2}) by introducing a new real-valued variable $z^i_{jk} = x_{jk} y_i$ and new linear constraints. Let $p_i^c$ and $w_i^c$ denote the total profit and the total weight of the compulsory items in city $i$ obtained with respect to Proposition~\ref{prop2}. Similarly, let $w_i^{max}$ be the total weight of the items (including all the compulsory items) in city $i$. Then variable $y_i$, $i=1, \ldots, n$, can be bounded from below by $L_i = 1 / \left(\upsilon_{max}- \nu\sum_{j=1}^i w_j^c\right)$ and from above by $U_i = 1 / \left(\upsilon_{max}- \nu \cdot min\left(\sum_{j=1}^i w_j^{max},W\right)\right)$. In summary, we can formulate {\NKPc} (and {\NKPu}) as the following linear mixed 0-1 program ({\ENKP}):

{\footnotesize
\begin{flalign}
\nonumber\mbox{max}\; &\text{{\ProfitMIP}} = \displaystyle\sum_{i=1}^n \left(p_i^c +\displaystyle\sum_{k=1}^{m_i} p_{ik} x_{ik} - R d_{i} y_i\right)
\\
\nonumber\mbox{s.t.}\; \, &\upsilon_{max}y_i + \nu \left(w_i^c + \displaystyle\sum_{j=1}^i \displaystyle\sum_{k=1}^{m_j} w_{jk} z^i_{jk} \right)= 1, \; i=1,\ldots,n
\\
\nonumber&z^i_{jk}\leq U_ix_{jk}, \; i,j=1,\ldots,n,\; j\leq i,\; e_{jk} \in M_{j}
\\
\nonumber&z^i_{jk}\geq L_ix_{jk}, \; i,j=1,\ldots,n,\; j\leq i,\; e_{jk} \in M_{j}
\\
\nonumber&z^i_{jk}\geq y_i + U_i\left(x_{jk}-1\right), \; i,j=1,\ldots,n, \; j\leq i,\; e_{jk} \in M_{j}
\\
\nonumber&z^i_{jk}\leq y_i + L_i\left(x_{jk}-1\right), \; i,j=1,\ldots,n, \; j\leq i,\; e_{jk} \in M_{j}
\\ 
&\displaystyle\sum_{i=1}^n \displaystyle\sum_{k=1}^{m_i} w_{ik} x_{ik} \leq W \label{eq:e6}
\\  
&\text{{\ProfitL}} \leq \text{{\ProfitMIP}} \leq \text{{\ProfitU}}\label{eq:e7}
\\  
\nonumber&x_{ik} \in \left\{0,1\right\}, \; e_{ik} \in M
\\ 
\nonumber&z^i_{jk}\in \mathbb{R}_+, \; i,j=1,\ldots,n, \; j\leq i,\; e_{jk} \in M_{j}
\\ 
\nonumber&y_i \in \mathbb{R}_+, \; i=1,\ldots,n
\end{flalign}}

A solution of {\ANKP} can be used as a starting solution for {\ENKP} and can yield the lower bound value {\ProfitL}. In its turn, {\ANKPu} can provide the upper bound value {\ProfitU}. To set the value of $\lambda$ to be used in both {\ANKP} and {\ANKPu}, we specify its value through the notation {\ENKP}. To strengthen the relaxation of {\ENKP}, the sequencing constraints of Section \ref{sec:constr} can be imposed as valid inequalities as has been earlier proposed in \cite{Polyakovskiy15}. However, our current investigations show that they are not beneficial anymore when the upper bound produced by {\ANKPu} is applied in Eq. \ref{eq:e7}. An effective set of inequalities in order to obtain tighter relaxations can be obtained from the reformulation-linearization technique (RLT) by~\cite{Sherali99}, which uses $3n$ additional inequalities for the capacity constraint (\ref{eq:e6}). Specifically, multiplying (\ref{eq:e6}) by $y_l$, $U_l-y_l$ and $y_l-L_l$, $l=1, \ldots, n$, we obtain the following inequalities:

{\footnotesize
\begin{flalign}
\nonumber&\displaystyle\sum_{i=1}^n \displaystyle\sum_{k=1}^{m_i} w_{ik} z^l_{ik} \leq Wy_l;
\\  
\nonumber&U_l\displaystyle\sum_{i=1}^n \displaystyle\sum_{k=1}^{m_i} w_{ik} x_{ik}- \displaystyle\sum_{i=1}^n \displaystyle\sum_{k=1}^{m_i} w_{ik} z^l_{ik} \leq U_lW - Wy_l;
\\ 
\nonumber&\displaystyle\sum_{i=1}^n \displaystyle\sum_{k=1}^{m_i} w_{ik} z^l_{ik} - L_l\displaystyle\sum_{i=1}^n \displaystyle\sum_{k=1}^{m_i} w_{ik} x_{ik} \leq Wy_l - L_lW.
\end{flalign}}

\section{Branch-Infer-and-Bound Approach} \label{sec:BIB}

Constraint programming (CP) has been shown to be a promising solution technique for various combinatorial optimization problems (\cite{Rossi:2006, Rossi2008}). It deals with a problem consisting of a set of variables $X=\left\{x_1,\ldots,x_n\right\}$ and a finite set of constraints $C$ given on the elements of $X$. Each variable $x_i \in X$ is associated with a domain $D_i$ of available values. When the domain of every variable $x_i$ is reduced to a singleton $\left\{v_i\right\}$, a values vector $v=\left(v_1,\ldots,v_n\right)$ is obtained. A satisfiability problem asks for a decision vector $x=v$ such that all the constraints in $C$ are satisfied simultaneously. A constraint optimization problem involves in addition an objective function $f\!\left(x\right)$ that is to be either maximized or minimized over the set of all feasible solutions. In CP, constraints are given in a declarative way, but are viewed individually as special-purpose procedures that operate on a solution space. Each procedure applies a filtering algorithm that eliminates those values from the domains of the involved variables which cannot be a part of any feasible solution with respect to that constraint. The restricted domains generated by the constraints are in effect elementary in-domain constraints that restrict a variable to a domain of possible values. They become a part of a constraint store. To link all the procedures together in order to solve a problem as a whole, the constraint store is passed on to the next constraint to be processed. In such a way, the results of one filtering procedure are propagated to the others. In general, filtering algorithms are called repeatedly to achieve a certain level of consistency. This is because achieving arc consistency for one constraint might make other constraints inconsistent. Specifically, constraint $c \in C$ involving variables $x_i$ and $x_j$ is said to be arc consistent with respect to $x_i$ if for each value $v' \in D_i$ there is an allowed value of $x_j$. A constraint satisfaction problem is arc consistent iff every constraint $c \in C$ is arc consistent with respect to $x_i$ as well as to $x_j$. Therefore, multiple runs of filtering are required for those constraints that share common variables of $X$. This process is called constraint propagation. CP aims to enumerate solutions with respect to the constraint store in order to find the best feasible solution. To cope with this, a search tree is used, and every variable $x_i$ with domain $D_i$ is examined in some node of the tree. If $D_i = \emptyset$, an infeasible solution is found. If $\left|D_i\right| >1$, one can branch on $x_i$ by partitioning $D_i$ into smaller domains, each corresponding to a branch. The domains of the variables decrease as they are reduced via constraint propagation when one descends into the tree. In the case of the constraint optimization problem, the search continues unless either the best solution is determined over those solutions where all the domains are singletons, or at least one of the domains is empty for every leaf node of the search tree. Certainly, the order in which the variables are instantiated and how the domains are partitioned matters for a running time.

Combining CP with the branch-and-cut method is a natural hybridization which results from the complementary strengths of both techniques. It gives rise to the so-called branch-infer-and-relax (BIR) approach presented by \cite{Bockmayr05}. The idea of BIR is to combine filtering and propagation used in CP with relaxation and cutting plane generation used in MIP. In each node of a search tree, constraint propagation creates a constraint store of in-domain constraints, while polyhedral relaxation creates a constraint store of inequalities. The two constraint stores can enrich each other, since reduced domains impose bounds on variables, and bounds on variables can reduce domains. The inequality relaxation is solved to obtain a bound on the optimal value, which prunes the search tree as in the branch-and-cut method. 

Here, to solve {\NKPc} (and {\NKPu}), we adopt this idea and introduce a branch-infer-and-bound approach that compounds CP and the upper bound introduced in Section \ref{sec:ub}. Specifically, we substitute the relaxation used in BIR with a stand-alone upper bound procedure to be executed in each node of the search tree in order to prune some of its branches. In each node, we create a refined set of items $M' = M \setminus \cup_{e_{ik} \in M} e_{ik}:\left|D_{ik}\right|=1$ and add the weight and profit of those items whose $D_{ik}=\left\{1\right\}$ to $w_i^c$ and $w_i^{max}$ in the model of {\ProfitU}, respectively. In other words, we treat the items accepted by the search as compulsory. Finally, we apply {\ANKPu} to $x$ formed on $M'$ and prune a branch if the resulting {\ProfitU} is smaller than the objective value of the best incumbent solution known.

Similarly to the previous MIP formulations, our CP model bases the search on binary decision vector $x$ where variable $x_{ik}$ takes the value of 1 to indicate that item $e_{ik} \in M$ is chosen. To speed up computations, it employes an auxiliary integer variable $w_i$ that calculates the total weight of the items selected in city $i$ and all the preceding cities when traveling along the edge $\left(i,i+1\right)$, for any $i=1,\ldots,n$. Again, $w_i^c$ and $p_i^c$ denote the total weight and the total profit of compulsory items collected in city $i$. Here, we assume that the both values come out of the pre-processing step. The model has the following objective function and constraints ({\PWTBIB}):

{\footnotesize
\begin{flalign}
\mbox{max} \;  &\text{{\ProfitBIB}} = \displaystyle\sum_{i=1}^n \left(p_i^c +\displaystyle\sum_{k=1}^{m_i} p_{ik} x_{ik} - \frac{Rd_i}{\upsilon_{max}-\nu w_i}\right) \label{eq:bib0}
\\
\mbox{s.t.} \;  &w_i=w_{i-1} + w_i^c+\displaystyle\sum_{e_{ik}\in M_i} w_{ik}x_{ik}, \; i=1,\ldots,n \label{eq:bib1}
\\
& \displaystyle\sum_{i=1}^n \left(w_i^c + \displaystyle\sum_{k=1}^{m_i} w_{ik} x_{ik}\right) \leq W \label{eq:bib3}
\\
&x_{il}\leq x_{ik}, \; i\!=\!1,\ldots,n, \, e_{il},e_{ik} \! \in M_i \,:\, \left(p_{il}<p_{ik}\right) \wedge \left(w_{il} \geq w_{ik}\right), \, l \neq k, \label{eq:bib4}
\\
&x_{jl}\leq x_{ik}, \; i\!=\!1,\ldots,n, \, j<i, \, e_{jl} \!\in\! M_j, \, e_{ik} \!\in\! M_i \!: \! \left(p_{jl}-\Delta_l^{ji}<p_{ik}\right) \!\wedge\! \left(w_{jl}\geq w_{ik}\right)\label{eq:bib5}
\\
&x_{jl}\geq x_{ik}, \; i\!=\!1,\ldots,n, \, j<i, \, e_{jl} \!\in\! M_j, \, e_{ik} \!\in\! M_i \!: \! \left(p_{jl}-\overline{\Delta}_l^{ji}>p_{ik}\right) \!\wedge\! \left(w_{jl}\leq w_{ik}\right)\label{eq:bib6}
\\
\nonumber&\text{\texttt{sequencing}}\!\left(x_{jl},\left[w_1,\ldots,w_n\right]\right), \; j=1,\ldots,n, 
\\
&\quad\quad\quad\quad\quad e_{jl}\in M_j \,: \, \left(\exists e_{ik} \in M_i, \, j<i\,: \, p_{jl}-\overline{\Delta}_l^{ji} \leq p_{ik} \leq p_{jl}-\Delta_l^{ji}\right)\label{eq:bib7}
\\ 
& x_{ik} \in \left\{0,1\right\}, \; e_{ik} \in M \label{eq:bib10}
\\
& w_i \in \left\{0,\ldots,W\right\}, \; i=1,\ldots,n \label{eq:bib11}
\\
& w_0 =0 \label{eq:bib12}
\end{flalign}
}

Function \ref{eq:bib0} represents the objective function of the problem. For each edge $\left(i,i+1\right)$, $i=1,\ldots,n$, it sums up the profits of items taken in city $i$ minus the cost of transportation of all the items that have been placed to the vehicle in city $i$ and all the cities prior to $i$. Equation (\ref{eq:bib1}) calculates the weight $w_i$ of all the items taken in the cities $1,\ldots,i$. Equation (\ref{eq:bib3}) is a capacity constraint. Equations (\ref{eq:bib4}), (\ref{eq:bib5}) and (\ref{eq:bib6}) impose the set of redundant sequencing constraints {\SCa}, {\SCb}, and {\SCc} of Section \ref{sec:constr}, respectively. This set may be rather small, and therefore might have a limited impact on inference of in-domain constraints during the search. Indeed, more constraints might be involved if $w_b^c$ in $\Delta_l^{ji}$ of constraint {\SCb} was larger and $w_b^{max}$ in $\overline{\Delta}_l^{ji}$ of constraint {\SCc} was smaller. The values of these two variables $w_b^c$ and $w_b^{max}$ can in fact be  considered as initial lower and upper bounds on the weight of the items collected in city $b$. As one descends into the tree, items are either collected or rejected. Being picked up in some city, an item contributes its weight that increases the lower bound. Being rejected, it lowers the upper bound. We include those constraints into the pool that still may work out when the lower or upper bound on the weight reaches a certain level. Specifically, (\ref{eq:bib7}) adds a redundant customized constraint for each item $e_{jl}$ that has at least one related item $e_{ik}$, $j<i$, such that their mutual sequencing depends on the weight of the items collected in cities $j,\ldots,i-1$. Equations (\ref{eq:bib10}) and (\ref{eq:bib11}) define the domains of the variables. Finally, Equation (\ref{eq:bib12}) sets the base case $w_0=0$. 

We assume a depth-first search strategy for traversing the binary search tree and instantiate variables in the order in which the cities appear in $N$. No order is given to the items within the same city. Therefore, at the moment when item $e_{ik}$ is to be instantiated by the search, the domains of the variables associated with the items in cities $1,\ldots,i-1$ and with those in city $i$ that appear prior to $e_{ik}$ in the set $M_i$ have been already reduced to singletons. Accordingly, the domain of variable $w_i$ is reduced to a singleton once the decision variables of the items in city $i$ have been all fixed to singletons. 

The customized sequencing constraint $\text{\texttt{sequencing}}\!\left(x_{jl},\left[w_1,\ldots,w_n\right]\right)$ applies to variable $x_{jl}$ for which there exists at least one item, say $x_{ik}$, such that $p_{jl}-\overline{\Delta}_l^{ji} \leq p_{ik} \leq p_{jl}-\Delta_l^{ji}$ holds and $j<i$. Algorithm \ref{alg:CC} sketches the pseudocode of the corresponding filtering algorithm. We use a Boolean variable $\Theta_{e_{jl}e_{ik}}$, which takes value $true$ to indicate the situation when selection of items $e_{jl}$ and $e_{ik}$ may be potentially sequenced. First, the algorithm initializes variables $\overline{w}^{max}$ and $\overline{w}^{min}$ that, respectively, represent the upper and lower bounds on the weight of collected items that the vehicle has in city $i$. It sets both variables to the sum of weights of items collected in the cities prior to city $j$, i.e. $w_{j-1}$, and adds the weight of item $e_{jl}$ if its variable $x_{jl}$ has been fixed to $1$ (cf. line \ref{q1}). Then the algorithm starts exploring the items positioned in the cities succeeding $j$. Each time the next city $i$ is taken into consideration, it adds the weight of all the items existing in $i$ to $\overline{w}^{max}$ and the weight of all the compulsory items in $i$ to $\overline{w}^{min}$ (cf. line \ref{q2}). In each city $i$, the algorithm examines the items that the city contains. When the corresponding variable $x_{ik}$ of item $e_{ik}$ is a singleton, the algorithm modifies the bounds on collected weight accordingly (cf. lines \ref{q3} and \ref{q4}). Subsequently, if $\Theta_{e_{jl}e_{ik}}$ is $true$, it tries to establish a sequencing relation between $e_{ik}$ and $e_{jl}$. If the condition in line \ref{q5} holds, it excludes 1 from domain $D_{ik}$, and therefore declines item $e_{ik}$ since $e_{jl}$ is assumed to be rejected by the search. At the same time, it lowers $\overline{w}^{max}$ by the value of $w_{ik}$. On the other hand, if the condition in line \ref{q6} holds, it excludes 0 from domain $D_{ik}$, and therefore selects item $e_{ik}$ because $e_{jl}$ is assumed to be selected. In addition, it increases $\overline{w}^{min}$ by the value of $w_{ik}$. If no relation has been established at that stage, the algorithm adds $e_{ik}$ to the set of items $M^*$. Continuing to explore other items, it tries to prove each item of $M^*$ to be unprofitable every time it finishes investigating city $i$ (cf. line \ref{q7}). Furthermore, having all the cities analyzed, the algorithm returns back to set $M^*$ and tries to prove each of its items to be compulsory if the instance $I$ at hands refers to {\NKPu} (cf. line \ref{q8}).

\SetAlFnt{\tiny}
\begin{algorithm}[!htb]\caption{The Filtering Algorithm of Constraint \newline $\text{\texttt{sequencing}}\!\left(x_{jl},\left[w_1,\ldots,w_n\right]\right)$}\label{alg:CC}
initialize $\overline{w}^{max} \leftarrow {w}_{j-1}+w_{jl}x_{jl}$; initialize $\overline{w}^{min} \leftarrow {w}_{j-1}+w_{jl}x_{jl}$\; \label{q1}
initialize $M^*\leftarrow \emptyset$\;
set $\overline{\Delta}_l^{jj} \leftarrow 0$; $\Delta_l^{jj} \leftarrow 0$\;
\For{each city $i$ from j to $n$}{
$\overline{w}^{max} \leftarrow \overline{w}^{max} + w_i^{max}$; $\overline{w}^{min} \leftarrow \overline{w}^{min} + w_i^c$\; \label{q2}
\For{each item $e_{ik} \in M_i, \, e_{ik} \neq e_{jl}$}{
initialize $flag \leftarrow true$\;
\If{\texttt{DomainSize}$\left(D_{ik}\right) \leq 1$}{
$flag \leftarrow false$\;
\lIf{$x_{ik}=0$}{$\overline{w}^{max} \leftarrow \overline{w}^{max}-w_{ik}$} \label{q3}
\lIf{$x_{ik}=1$}{$\overline{w}^{min} \leftarrow \overline{w}^{min}+w_{ik}$} \label{q4}
}
\If{$\Theta_{e_{jl}e_{ik}}$}{
\If{$\left(w_{jl} \leq w_{ik}\right) \wedge \left(p_{jl}-\overline{\Delta}_l^{ji} > p_{ik}\right) \wedge \left(x_{jl}=0\right)$}{ \label{q5}
\If{\texttt{DomainSize}$\left(D_{ik}\right) = 2$}{
$\overline{w}^{max} \leftarrow \overline{w}^{max}-w_{ik}$\;
$flag \leftarrow false$\;
}
\texttt{RemoveValue}$\left(D_{ik},1\right)$\;
}
\If{$\left(w_{jl} \geq w_{ik}\right) \wedge \left(p_{jl}-\Delta_l^{ji} < p_{ik}\right) \wedge \left(x_{jl}=1\right)$}{ \label{q6}
\If{\texttt{DomainSize}$\left(D_{ik}\right) = 2$}{
$\overline{w}^{min} \leftarrow \overline{w}^{min}+w_{ik}$\;
$flag \leftarrow false$\;
}
\texttt{RemoveValue}$\left(D_{ik},0\right)$\;
}
}
\If{$flag$}{
$M^*\leftarrow M^*\cup \left\{e_{ik}\right\}$\;
initialize $\overline{\Delta}_k^{in+1} \leftarrow 0$; $\Delta_k^{in+1} \leftarrow 0$\;
}
}
$\overline{\Delta}_l^{ji+1} \leftarrow \overline{\Delta}_l^{ji} + Rd_i\left(\frac{1}{\upsilon_{max}-\nu\min\left(\overline{w}^{max},W\right)}-\frac{1}{\upsilon_{max}-\nu\left(\min\left(\overline{w}^{max},W\right)-w_{ik}\right)}\right)$\;
$\Delta_l^{ji+1} \leftarrow \Delta_l^{ji} +Rd_j\left(\frac{1}{\upsilon_{max}-\nu\left(\min\left(\overline{w}^{min},W\right)+w_{ik}\right)}-\frac{1}{\upsilon_{max}-\nu\min\left(\overline{w}^{min},W\right)}\right)$\;
\For{each item $e_{ab} \in M^*$}{
$\overline{\Delta}_b^{an+1} \leftarrow \overline{\Delta}_b^{an+1} + Rd_i\left(\frac{1}{\upsilon_{max}-\nu\min\left(\overline{w}_j^{max},W\right)}-\frac{1}{\upsilon_{max}-\nu\left(\min\left(\overline{w}^{max},W\right)-w_{ab}\right)}\right)$\;
$\Delta_b^{an+1} \leftarrow \Delta_b^{an+1} +Rd_j\left(\frac{1}{\upsilon_{max}-\nu\left(\min\left(\overline{w}^{min},W\right)+w_{ab}\right)}-\frac{1}{\upsilon_{max}-\nu\min\left(\overline{w}^{min},W\right)}\right)$\;
\If{$p_{ab}-\Delta_b^{an+1} \leq 0$}{ \label{q7}
\texttt{RemoveValue}$\left(D_{ab},1\right)$\;
$\overline{w}^{max} \leftarrow \overline{w}^{max}-w_{ab}$\;
$M^*\leftarrow M^*\setminus \left\{e_{ab}\right\}$\;
}
}
}
\If{\texttt{ProblemType}$\left(I\right) \equiv $\NKPu}{
\For{each item $e_{ab} \in M^*$}{
\lIf{$p_{ab}-\overline{\Delta}_b^{an+1} > 0$}{\texttt{RemoveValue}$\left(D_{ab},0\right)$}\label{q8}
}
}
\end{algorithm}

\section{Computational Experiments} \label{sec:CE}

In this section, we investigate the effectiveness of the proposed approaches by experimental studies. On the one hand, we assess the advantage of the pre-processing scheme in terms of quantity of discarded items and auxiliary decision variables. On the other hand, we evaluate our {\ANKP}, {\ENKP}, and {\PWTBIB} models in terms of solution quality and running time. The program code is implemented in JAVA using the \textsc{IBM Optimization Studio} 12.6.2. To solve the mixed-integer programs {\ANKP}, {\ANKPu} and {\ENKP}, we use \textsc{Cplex} with default settings. When running {\ANKPu}, we increase \textsc{Cplex}'s precision by setting the relative tolerance on the gap between the best integer objective and the objective of the best node remaining to $1e\!-\!7$. To solve the constrained program within {\PWTBIB}, we use \textsc{CP Optimizer} switched to the depth-first search mode and set the relative tolerance gap to 0. Furthermore, we limit the parallel mode to only a single thread for \textsc{Cplex} and \textsc{CP Optimizer} to make them both comparable to each other when dealing with the small size instances in Section~\ref{sec:ss}. We set the number of threads to the maximum number of cores available when investigating the large size instances in Section~\ref{sec:ls}. 

The test instances are adopted from the benchmark set $B$ of \cite{Polyakovskiy14}. This benchmark set is constructed on TSP instances from TSPLIB introduced by \cite{Reinelt91} augmented by a set of items distributed among all the cities but the first one. We use the set of items available in each city and obtain the route from the corresponding TSP instance by running the Chained Lin-Kernighan heuristic proposed by \cite{chainedLK03applegate}. Given the permutation $\pi = (\pi_1, \pi_2, \ldots, \pi_n)$ of the cities computed by the Chained Lin-Kernighan heuristic, where $\pi_1$ is free of items, we use $N=(\pi_2, \pi_3, \ldots, \pi_n, \pi_1)$ as the route for our problem. We consider the \textit{uncorrelated} ({\U}), \textit{uncorrelated with similar weights} ({\USW}), and \textit{bounded strongly correlated} ({\BSC}) types of items' generation, and set $\upsilon_{min}$ and $\upsilon_{max}$ to 0.1 and 1 as proposed for $B$.

\subsection{Computational Experiments on the Set of Small Size Instances} \label{sec:ss}

Here, using a set of small instances, our goal is to evaluate the performance of our pre-processing scheme and the performance of the proposed approximate and exact approaches. Unlike experiments carried out in our earlier research (\cite{Polyakovskiy15}), here we are able to find optimal solutions to all the small instances within the same time limit. Therefore, the main focus of our investigation with respect to the exact approaches is their running times rather than any qualitative performance measures.

We study three families of small size instances based on the TSP problems \texttt{eil51}, \texttt{eil76}, and \texttt{eil101} with 51, 76 and 101 cities, respectively. This series of experiments has been carried out on PC with 4 Gb RAM and a 3.06 GHz Dual Core processor. The results of the experiments are shown in Table \ref{tab:res1}. All the instances of a family have the same route $N$. We consider instances with 1, 5, and 10 items per city. The postfixes 1, 6 and 10 in the instances' names indicate the vehicle's capacity $W$. The greater the value of a postfix is, the larger $W$ is given. Column 2 specifies the total number of items $m$. Ratio $\alpha=100 \cdot \left(m-m'\right)/m$ in Column $3$ denotes a percentage of items discarded in a pre-processing step, where $m'$ is the number of items left after pre-processing. Column $ver$ identifies by \textit{``u''} whether {\NKPc} has been reduced to {\NKPu} by pre-processing. Columns 5-7 report results for {\ANKP} with $\lambda=100$. Specifically, column 5 gives $\rho$ as a ratio between the lower bound obtained by {\ANKPone} and the optimum obtained by the branch-infer-and-bound approach. Column 6 contains the running time $t$ of {\ANKPone}. Column 7 shows a rate $\beta$ that is a percentage of auxiliary $y$-type variables used in practice by {\ANKP}. At most $\lambda n$ variables is required by {\ANKP}. Thus, $\beta$ is computed as $\beta=100 \cdot \left(\sum_{i=1}^n \left|B_i\right|\right)/\left(\lambda n\right)$. Column 8 presents the running time $t$ for the MIP-based exact approach {\ENKP} when $\lambda$ is set to 1000. Therefore, both {\ANKP} and {\ANKPu} incorporated into {\ENKP} use $\lambda=1000$ as well to compute initial lower and upper bounds. The time limit of 1 day has been given to {\ENKP} in total, while {\ANKP} and {\ANKPu} have got the time limit of 2 hours each. The running time of {\ENKP} provided in the table includes the total time taken by {\ANKP} and {\ANKPu}. For most of the instances except the instance ``\texttt{uncorr-s-w\_01}'' of the family \texttt{eil101}, this time is found negligible. Column 9 reports $\omega$ as a relative gap in percents that compares the running time of {\ENKP} to the smallest running time over all the algorithms studied in the experiment. In general, $\omega$ is to be computed as $\omega=100 \cdot \left(t^{A\!L\!G}-t^{M\!I\!N}\right)/t^{M\!I\!N}$, where $t^{A\!L\!G}$ is the running time of a particular algorithm and $t^{M\!I\!N}$ is the minimum over the running times of the various configurations of {\ENKP} and {\PWTBIB} investigated here.

The rest columns of the table describe results for the branch-infer-and-bound approach {\PWTBIB} with $\lambda \in \left\{500, 1000, 1500\right\}$. Furthermore, two cases of {\PWTBIB} for $\lambda=1000$ have been studied. {\PWTBIBtwo} is exactly that one which is described in Section \ref{sec:BIB}. {\PWTBIBthree} is its copy that does not include the customized sequencing constraints to the model. In such way, we evaluate the impact of the constraint on the performance of {\PWTBIB}. We employ {\ANKPone}, which we give 2 hours of running time limit, to provide {\PWTBIB} with a lower bound to be used then to prune the search tree. Within {\PWTBIBone}, {\PWTBIBtwo}, {\PWTBIBthree}, and {\PWTBIBfour}, we run the variant of {\ANKPu} where the integrality constraints on $x$-type decision variables are removed. This makes {\ANKPu} a linear program and significantly speeds up computations at the very small cost of solution quality. Each of the columns $t$ reports the total computational time for the corresponding {\PWTBIB} and includes the time taken by {\ANKPone}. Similarly, each of the columns named $\omega$ does when reporting $\omega$ that compares the running time of {\PWTBIB} to the smallest running time found over the studied {\ENKP} and {\PWTBIB} approaches. The least running time obtained for a particular instance is marked by bold and underlined in the entry $t$ of the corresponding approach. The entries of the table marked by ``-'' indicate that optimal solutions have not been obtained within the given time limit. 

\subsubsection{Performance of the Pre-processing Scheme} \label{sec:sss1}

Here, we evaluate the performance of our pre-processing scheme by calculating the percentage of discarded items and auxiliary decision variables. We aim to understand for which classes and types of the instances the pre-processing scheme works fine and which instances are hard to be reduced. Furthermore, we wonder how many of the instances of {\NKPc} become those of {\NKPu}.

The results of the experiments demonstrate efficiency of the pre-processing scheme. It is rather good with respect to the instances of {\U} type and removes on average 31.6\% of their items. Concerning the {\USW} type of the instances, it is able to exclude on average 18.5\% of the items that they contain. Within these two categories, the instances with large $W$ are rather liable to reduction to instances of {\NKPu}. Because $W$ is large, they get more chances to loose enough items so that the total weight of rest items becomes less or equal to $W$. The pre-processing scheme does not work well for the {\BSC} type of the instances. No instance of this type has been reduced to {\NKPu}. Because the profit of an item approaches its weight, the pre-processing encounters a difficulty to find unprofitable items for this instance type. In general, the way of how profits and weights are generated is not an obstacle in itself for the pre-processing to be successful. There are other factors, like the value of rent rate $R$, the value of capacity $W$, and a distance to the destination from the city where an item is positioned, that hinder its application. For example, if a route is long enough, some items in the first cities can be shown to be unprofitable even in the case of their {\BSC} type of generation. Clearly, the fact that we cannot handle the instances of this type is a proper property of the benchmark suite $B$. 

Obviously, discarding items within pre-processing reduces the number of auxiliary variables in {\ANKP}. The rate $\beta$ demonstrates that in practice {\ANKPone} uses a very reduced set of them. The average over all the entries is just 48.5\%. Therefore, less than a half of all possible variables is used only. In general, $\beta$ is significantly small when $W$ is large, since latter results in a slower growth of diapason $\left[\upsilon_i^{\min},\upsilon_i^{\max}\right]$ in {\ANKP}, for $i=1,\ldots,n$. In other words, the instances with large $W$ require less number of auxiliary decision variables comparing to the instances where $W$ is smaller.

\subsubsection{Performance of the Approximate Approach} \label{sec:sss2}

We now aim to evaluate the performance of {\ANKPone} concerning its running time and solution quality compared to optima. {\ANKPone} is particularly fast and its model is solved to optimality in a very short time for all the small size instances. Only one instance of the whole test suite causes a difficulty in terms of running time. The approximate approach looks very swift even with instances of the {\BSC} type and produces very good approximation for reasonably small $\lambda=100$. The ratio $\rho$ close to $1$ points out that {\ANKPone} obtains approximately the same result as the optimal objective value is, but in a shorter time. Therefore, {\ANKP} gives an advanced trade-off in terms of computational time and solution's quality comparing to the exact approaches. The larger $\lambda=1000$ has been tested in our earlier experiments (\cite{Polyakovskiy15}). However, it results to a very limited improvement in the value of the total reward at the larger cost of running time, and more importantly at the larger cost of memory consumption. Indeed, as $\lambda$ increases, the approach requires more memory as the number of auxiliary variables grows.

\subsubsection{Performance of the Exact Approaches} \label{sec:sss3}

In this part of the analysis, our goal is to evaluate the performance of the {\ENKP} and {\PWTBIB} approaches and determine which classes and types of the instances are hard to be solved by each of them. Since all the instances can be solved to optimality, we treat a time spent to achieve an optimal solution as hardness of an instance.

Comparing to the earlier results, we see now that the unconstrained instances of the problem are not to be easier to handle as it has been previously observed (\cite{Polyakovskiy14,Polyakovskiy15}). Now, our mixed-integer programming approach is able to solve all the small size instances to optimality as it is strengthened with the upper bound {\ANKPu}. Specifically, {\ENKP} takes much less time than the given limit and can be executed on an ordinary PC instead of the highly productive computational cluster that has been previously utilized. 

The results show that the instances of the {\USW} type are harder to solve for {\ENKP} comparing to other types. This fact looks interesting. Comparing to the {\BSC} type, this type has around 20\% of items per instance excluded by the pre-processing step and needs much less auxiliary variables, but takes more time to achieve an optimal solution. This also differs {\PWT} from the classical 0-1 knapsack problem for which the {\BSC} type is shown to be the hardest one (\cite{Martello99}). The {\BSC} type is not easier to solve than the {\U} type in terms of running time, nor the latter is when compared to the former. 

{\ENKP} outperforms {\PWTBIB} mainly on the set of {\BSC} type instances with the least capacity $W$. {\PWTBIBone} is superior on the wide range of instances. It is highly effective for the set of {\U} and {\USW} type instances with large capacity. Depending on the parameter $\lambda$, performance of {\PWTBIB} can be further improved for some instances. Specifically, setting $\lambda$ to larger values allows {\PWTBIB} to solve {\BSC} type instances with many items and large capacity faster. The further increase of $\lambda$ up to 1500, leads to the best result for some of those instances. However, this degrades performance of {\PWTBIB} on other instances. Performance of {\PWTBIB} with the values of $\lambda$ less than 500 and greater than 1500 have also been investigated. However, using too small or too large values increases the running time of the approach. The same behavior has been observed for {\ENKP}, for which the value of 1000 is the most promising one. In summary, we argue that selecting $\lambda=1000$ represents a good balance when the classification of instances is unknown, e.g. a new problem instance is to be solved. {\PWTBIBtwo} works reasonably fast over all the instances. Although it loses against {\PWTBIBone} on many instances, these loss are insignificant when compared to gains on some other instances, for example, on those ones prefixed by ``\texttt{b-s-corr\_10}''.

\subsubsection{Impact of the Sequencing Constraints} \label{sec:sss4}

Here, we are interested in understanding the impact of the sequencing constraints on the performance of the CP Search. We wonder how strong the constraint can be in pruning the search tree and for which types and classes of the instances it performs well.

Table \ref{tab:Ares1} presents the details on the constraint programming search performed by {\PWTBIB} on small size instances of the benchmark suite. Columns 1-3 specify the instance's name, the total number of items $m$, and the version of the problem solved. Other columns of the table describe results for {\PWTBIB} with $\lambda \in \left\{500, 1000, 1500\right\}$. Each of the sections shows the number of branches $b$ totally explored and the number of fails $f$ obtained by the CP solver during the search when the corresponding parameter value $\lambda$ is in use. Furthermore, each of the columns ``\textit{{\ANKPu} runs}'' reports the number of successful {\ANKPu} runs, say $r^s$, that resulted in pruning of the search tree and the total number of runs, say $r^t$. The two values are separated by ``$\mid$''. In the parentheses, it also gives a percentage of successful runs $\eta$ computed as $\eta=100 \cdot r^s/r^t$.

The sequencing constraints of {\PWTBIB} look weak when dealing with the instances of the {\BSC} type. This can be observed from the results, which show that the number of {\ANKPu} runs considerably decreases when $\lambda$ increases. This means that {\PWTBIB} needs to tighten the upper bound by increasing the value of $\lambda$ in order to achieve a better performance for this type of the instances. In other words, {\PWTBIB} relies more on a tight upper bound to prune the search tree rather than on the sequencing constraints. In contrast, considering other types of the instances, the growth of the number of runs is not that much for them. Moreover, the number of explored branches $b$ and the number of fails $f$ obtained remain almost the same for different values of $\lambda$. This means that {\PWTBIB} extensively rely on the sequencing constraints when solving the {\U} and {\USW} types of the instances. When we turn off the sequencing constraints in the case of {\PWTBIBthree}, the approach explores more branches with respect to the {\U} and {\USW} types and spends more time to do this, while the number of branches traversed and the running time stay almost the same for the {\BSC} type. The reason why the sequencing constraints are weak with respect to the {\BSC} type is the same as for the pre-processing scheme which also has low performance when dealing with this type.

\begin{table}[!htbp]
\centering
\caption{Results of Computational Experiments on Small Size Instances}
\label{tab:res1}
{\tiny
\begin{adjustwidth}{-1cm}{}
\begin{tabular}{
@{\,}r@{\,}||
@{\,}r@{\,}|
@{\,}r@{\,}|
@{\,}c@{\,}||
@{\,}r@{\,}|@{\,}r@{\,}|@{\,}r@{\,}
@{}r@{}||
@{\,}r@{\,}|@{\,}r@{\,}
@{}r@{}||
@{\,}r@{\,}|@{\,}r@{\,}
@{}r@{}||
@{\,}r@{\,}|@{\,}r@{\,}
@{}r@{}||
@{\,}r@{\,}|@{\,}r@{\,}
@{}r@{}||
@{\,}r@{\,}|@{\,}r@{\,}
}
\hline
\multirow{2}{*}{instance}&\multirow{2}{*}{m}&\multirow{2}{*}{$\alpha$, \%}&\multirow{2}{*}{{\textit{ver}}}&\multicolumn{3}{c}{{\ANKPone}}&&\multicolumn{2}{c}{\ENKPone}&&\multicolumn{2}{c}{\PWTBIBone}&&\multicolumn{2}{c}{\PWTBIBtwo}&&\multicolumn{2}{c}{\PWTBIBthree}&&\multicolumn{2}{c}{\PWTBIBfour}\\
\hhline{~~~~------------------}
&&&& $\rho$ & $t$, sec & $\beta$, \% 
&&$t$, sec & $\omega$, \%
&&$t$, sec & $\omega$, \%
&&$t$, sec & $\omega$, \% 
&&$t$, sec & $\omega$, \% 
&&$t$, sec & $\omega$, \% \\
\hline
\hline
\multicolumn{22}{c}{instance family \texttt{eil51}} \\
\hline
uncorr\_01&50&42.0&c&1.00000&0.2&55.8&&2.3&34.9&&\textbf{\underline{1.7}}&0.0&&3.5&107.4&&6.4&278.1&&6.1&256.9\\
uncorr\_06&50&14.0&c&1.00000&0.2&39.2&&3.2&255.5&&\textbf{\underline{0.9}}&0.0&&1.7&82.8&&3.2&252.2&&2.6&184.7\\
uncorr\_10&50&12.0&u&1.00000&0.1&11.1&&0.8&104.2&&\textbf{\underline{0.4}}&0.0&&0.5&30.9&&0.6&50.2&&0.6&58.0\\
uncorr-s-w\_01&50&30.0&c&1.00000&0.3&77.4&&2.9&84.3&&\textbf{\underline{1.6}}&0.0&&3.1&94.1&&7.6&381.5&&4.8&205.0\\
uncorr-s-w\_06&50&24.0&c&1.00000&0.1&35.8&&1.4&52.6&&\textbf{\underline{0.9}}&0.0&&1.7&84.8&&2.3&144.7&&2.7&190.9\\
uncorr-s-w\_10&50&34.0&u&1.00000&0.1&13.2&&1.6&379.2&&\textbf{\underline{0.3}}&0.0&&0.4&9.6&&0.5&36.1&&0.4&28.1\\
b-s-corr\_01&50&4.0&c&1.00000&0.3&89.7&&\textbf{\underline{4.5}}&0.0&&23.2&412.4&&49.0&982.1&&52.0&1049.0&&77.8&1620.7\\
b-s-corr\_06&50&0.0&c&1.00000&0.2&53.4&&\textbf{\underline{2.5}}&0.0&&2.9&14.0&&6.4&152.1&&6.3&146.0&&11.0&333.1\\
b-s-corr\_10&50&0.0&c&1.00000&0.2&25.7&&\textbf{\underline{1.9}}&0.0&&1.9&0.9&&3.7&100.1&&3.5&89.3&&6.4&241.0\\
uncorr\_01&250&39.2&c&1.00000&0.3&65.5&&10.4&127.5&&\textbf{\underline{4.6}}&0.0&&9.1&97.6&&21.5&368.7&&14.7&220.7\\
uncorr\_06&250&16.4&c&1.00000&0.2&38.3&&65.1&2353.2&&\textbf{\underline{2.7}}&0.0&&4.2&57.0&&7.3&175.2&&6.1&131.7\\
uncorr\_10&250&54.4&u&1.00000&0.1&10.9&&20.4&2512.1&&\textbf{\underline{0.8}}&0.0&&1.0&28.1&&1.8&131.3&&1.2&60.0\\
uncorr-s-w\_01&250&20.8&c&1.00000&0.3&88.0&&7.1&97.4&&\textbf{\underline{3.6}}&0.0&&6.8&87.9&&40.8&1026.9&&10.4&187.1\\
uncorr-s-w\_06&250&14.0&c&1.00000&0.2&44.6&&41.7&2520.9&&\textbf{\underline{1.6}}&0.0&&2.4&50.0&&5.8&262.0&&3.5&118.8\\
uncorr-s-w\_10&250&19.2&u&0.99998&0.2&15.7&&83.3&7761.7&&\textbf{\underline{1.1}}&0.0&&1.3&19.2&&2.3&120.0&&1.5&42.7\\
b-s-corr\_01&250&0.0&c&1.00000&0.3&90.2&&\textbf{\underline{8.9}}&0.0&&28.9&223.5&&58.3&552.7&&65.6&634.2&&94.1&953.0\\
b-s-corr\_06&250&0.0&c&0.99997&0.2&55.8&&\textbf{\underline{20.9}}&0.0&&27.3&31.0&&53.7&157.3&&55.4&165.7&&57.4&175.0\\
b-s-corr\_10&250&0.0&c&1.00000&0.2&26.7&&55.6&633.3&&\textbf{\underline{7.6}}&0.0&&9.8&29.4&&9.5&25.5&&16.1&112.5\\
uncorr\_01&500&37.0&c&1.00000&0.3&67.7&&12.9&12.0&&\textbf{\underline{11.5}}&0.0&&22.1&91.4&&68.5&493.1&&35.8&210.4\\
uncorr\_06&500&15.2&c&0.99993&0.3&38.8&&81.6&1369.8&&\textbf{\underline{5.6}}&0.0&&8.3&49.2&&17.8&221.4&&12.0&116.6\\
uncorr\_10&500&51.4&u&1.00000&0.2&11.6&&93.1&7093.6&&\textbf{\underline{1.3}}&0.0&&1.5&19.4&&3.4&160.2&&1.9&43.4\\
uncorr-s-w\_01&500&20.2&c&1.00000&0.2&89.1&&12.1&149.1&&\textbf{\underline{4.9}}&0.0&&8.7&79.2&&63.3&1200.5&&13.1&168.9\\
uncorr-s-w\_06&500&15.2&c&0.99990&0.2&44.2&&147.7&4854.8&&\textbf{\underline{3.0}}&0.0&&3.9&32.4&&9.9&231.5&&5.3&78.5\\
uncorr-s-w\_10&500&18.6&u&1.00000&0.2&16.1&&208.2&9788.2&&\textbf{\underline{2.1}}&0.0&&2.3&11.1&&4.6&118.6&&2.6&21.5\\
b-s-corr\_01&500&0.0&c&0.99993&0.3&91.3&&\textbf{\underline{29.9}}&0.0&&226.7&657.2&&324.1&982.6&&396.9&1225.8&&535.0&1687.3\\
b-s-corr\_06&500&0.0&c&0.99995&0.3&55.4&&\textbf{\underline{71.3}}&0.0&&316.7&344.1&&149.0&109.0&&161.5&126.4&&235.7&230.4\\
b-s-corr\_10&500&0.0&c&1.00000&0.2&26.0&&100.8&294.4&&87.9&243.8&&25.7&0.4&&\textbf{\underline{25.6}}&0.0&&27.4&7.0\\
\hline
\hline
\multicolumn{22}{c}{instance family \texttt{eil76}} \\
\hline
uncorr\_01&75&26.7&c&1.00000&0.3&76.7&&5.4&9.5&&\textbf{\underline{5.0}}&0.0&&10.4&110.1&&20.7&316.6&&17.7&257.6\\
uncorr\_06&75&14.7&c&1.00000&0.3&33.9&&9.8&433.3&&\textbf{\underline{1.8}}&0.0&&3.3&80.5&&4.3&134.7&&5.2&183.2\\
uncorr\_10&75&48.0&u&1.00000&0.1&11.3&&1.9&303.0&&\textbf{\underline{0.5}}&0.0&&0.6&20.1&&1.0&112.2&&0.7&53.2\\
uncorr-s-w\_01&75&26.7&c&1.00000&0.4&78.2&&4.9&2.3&&\textbf{\underline{4.8}}&0.0&&9.6&99.0&&30.4&529.9&&15.6&224.3\\
uncorr-s-w\_06&75&17.3&c&1.00000&0.3&40.7&&7.8&495.3&&\textbf{\underline{1.3}}&0.0&&2.2&67.9&&5.0&283.8&&3.2&147.7\\
uncorr-s-w\_10&75&16.0&u&1.00000&0.2&16.6&&10.8&1437.9&&\textbf{\underline{0.7}}&0.0&&0.9&31.1&&1.5&119.4&&1.1&59.8\\
b-s-corr\_01&75&0.0&c&1.00000&0.4&93.5&&\textbf{\underline{6.2}}&0.0&&215.6&3371.7&&463.3&7362.1&&510.3&8119.1&&770.3&12305.6\\
b-s-corr\_06&75&0.0&c&1.00000&0.3&58.9&&8.6&68.3&&\textbf{\underline{5.1}}&0.0&&11.1&117.7&&11.5&124.7&&19.1&272.8\\
b-s-corr\_10&75&0.0&c&1.00000&0.3&25.5&&6.7&15.5&&\textbf{\underline{5.8}}&0.0&&9.8&68.7&&10.1&73.8&&11.1&90.4\\
uncorr\_01&375&38.1&c&1.00000&0.3&66.3&&30.6&204.5&&\textbf{\underline{10.0}}&0.0&&19.9&98.0&&48.9&386.5&&32.4&222.2\\
uncorr\_06&375&16.0&c&1.00000&0.2&37.0&&162.0&2825.0&&\textbf{\underline{5.5}}&0.0&&9.6&73.0&&17.5&216.1&&14.5&162.6\\
uncorr\_10&375&9.9&u&1.00000&0.2&11.8&&105.4&7412.1&&\textbf{\underline{1.4}}&0.0&&1.6&15.6&&6.0&329.1&&1.8&30.0\\
uncorr-s-w\_01&375&14.9&c&1.00000&1.0&89.7&&26.4&205.8&&\textbf{\underline{8.6}}&0.0&&15.3&77.1&&347.2&3917.0&&24.3&181.0\\
uncorr-s-w\_06&375&12.3&c&1.00000&0.3&46.8&&165.9&4625.3&&\textbf{\underline{3.5}}&0.0&&5.4&52.5&&15.7&348.4&&7.4&110.7\\
uncorr-s-w\_10&375&14.9&u&1.00000&0.2&17.0&&230.9&10782.5&&\textbf{\underline{2.1}}&0.0&&2.4&14.7&&4.1&94.9&&2.9&34.5\\
b-s-corr\_01&375&0.0&c&1.00000&0.3&94.1&&\textbf{\underline{24.4}}&0.0&&51.2&110.4&&100.2&311.6&&116.8&379.6&&154.4&533.9\\
b-s-corr\_06&375&0.0&c&1.00000&0.3&56.6&&83.5&47.0&&149.0&162.1&&\textbf{\underline{56.8}}&0.0&&59.4&4.6&&98.2&72.9\\
b-s-corr\_10&375&0.0&c&0.99998&0.3&27.5&&181.4&579.8&&92.7&247.3&&\textbf{\underline{26.7}}&0.0&&27.6&3.5&&37.8&41.8\\
uncorr\_01&750&32.5&c&1.00000&0.4&71.4&&92.1&332.9&&\textbf{\underline{21.3}}&0.0&&33.4&57.0&&131.9&520.2&&52.8&148.3\\
uncorr\_06&750&14.8&c&1.00000&0.2&39.0&&429.8&2564.3&&19.6&21.7&&\textbf{\underline{16.1}}&0.0&&40.0&148.1&&22.8&41.1\\
uncorr\_10&750&43.1&u&1.00000&0.2&13.0&&306.7&8792.8&&3.8&10.4&&\textbf{\underline{3.4}}&0.0&&10.4&201.6&&3.6&3.6\\
uncorr-s-w\_01&750&16.7&c&1.00000&0.5&88.7&&117.0&950.9&&\textbf{\underline{11.1}}&0.0&&19.2&72.0&&255.2&2190.9&&28.9&159.8\\
uncorr-s-w\_06&750&13.5&c&1.00000&0.2&45.7&&472.0&7119.9&&\textbf{\underline{6.5}}&0.0&&7.8&19.6&&23.5&258.9&&10.2&56.3\\
uncorr-s-w\_10&750&14.4&u&1.00000&0.3&17.0&&823.7&17214.5&&\textbf{\underline{4.8}}&0.0&&5.1&7.3&&10.8&127.2&&5.6&17.7\\
b-s-corr\_01&750&0.0&c&0.99999&0.3&93.7&&\textbf{\underline{161.4}}&0.0&&183.8&13.8&&287.4&78.0&&388.3&140.5&&442.4&174.0\\
b-s-corr\_06&750&0.0&c&1.00000&0.5&55.4&&259.6&67.3&&975.0&528.1&&175.8&13.2&&203.1&30.9&&\textbf{\underline{155.2}}&0.0\\
b-s-corr\_10&750&0.0&c&0.99999&0.2&25.9&&281.7&106.7&&20261.7&14767.4&&176.7&29.7&&185.2&35.9&&\textbf{\underline{136.3}}&0.0\\
\hline
\hline
\multicolumn{22}{c}{instance family \texttt{eil101}} \\
\hline
uncorr\_01&100&49.0&c&1.00000&0.4&60.7&&7.4&144.7&&\textbf{\underline{3.0}}&0.0&&6.0&98.5&&12.9&325.4&&10.3&237.6\\
uncorr\_06&100&16.0&c&0.99993&0.3&39.7&&20.5&746.2&&\textbf{\underline{2.4}}&0.0&&4.0&65.9&&8.3&241.5&&6.0&147.2\\
uncorr\_10&100&57.0&u&1.00000&0.2&10.0&&4.8&524.7&&\textbf{\underline{0.8}}&0.0&&1.0&28.5&&1.8&136.2&&1.2&63.4\\
uncorr-s-w\_01&100&25.0&c&1.00000&0.3&90.3&&6.8&70.2&&\textbf{\underline{4.0}}&0.0&&7.9&95.3&&25.7&537.4&&12.4&207.0\\
uncorr-s-w\_06&100&17.0&c&1.00000&0.4&41.9&&17.2&1001.7&&\textbf{\underline{1.6}}&0.0&&2.6&65.0&&6.0&285.0&&3.8&144.2\\
uncorr-s-w\_10&100&15.0&u&1.00000&0.2&17.2&&39.6&3748.5&&\textbf{\underline{1.0}}&0.0&&1.3&25.6&&2.1&106.7&&1.6&59.1\\
b-s-corr\_01&100&0.0&c&1.00000&0.5&94.4&&\textbf{\underline{12.7}}&0.0&&60.2&373.0&&126.5&893.6&&135.5&964.3&&209.2&1543.2\\
b-s-corr\_06&100&0.0&c&1.00000&0.4&56.2&&11.3&8.5&&\textbf{\underline{10.4}}&0.0&&22.9&119.6&&22.9&119.1&&40.1&284.5\\
b-s-corr\_10&100&0.0&c&0.99990&0.2&28.2&&16.6&4.5&&\textbf{\underline{15.9}}&0.0&&27.3&71.7&&27.0&69.3&&41.6&161.3\\
uncorr\_01&500&38.8&c&1.00000&0.4&65.9&&31.6&120.2&&\textbf{\underline{14.4}}&0.0&&28.0&94.8&&76.3&430.7&&43.8&204.8\\
uncorr\_06&500&14.4&c&1.00000&0.3&39.2&&397.3&4678.5&&\textbf{\underline{8.3}}&0.0&&13.7&64.2&&26.3&216.9&&19.6&135.7\\
uncorr\_10&500&51.4&u&1.00000&0.2&11.4&&212.8&9707.2&&\textbf{\underline{2.2}}&0.0&&2.5&16.5&&5.6&157.1&&3.0&38.8\\
uncorr-s-w\_01&500&20.4&c&1.00000&4.5&88.4&&88.7&293.4&&\textbf{\underline{22.5}}&0.0&&39.4&74.7&&1924.5&8437.4&&60.2&167.2\\
uncorr-s-w\_06&500&14.2&c&1.00000&0.4&44.8&&365.2&6825.2&&\textbf{\underline{5.3}}&0.0&&7.7&46.2&&20.7&293.1&&10.7&102.7\\
uncorr-s-w\_10&500&16.4&u&1.00000&0.2&16.3&&525.4&16328.3&&\textbf{\underline{3.2}}&0.0&&3.5&10.7&&7.3&128.3&&4.0&25.4\\
b-s-corr\_01&500&0.0&c&1.00000&0.4&93.5&&\textbf{\underline{64.6}}&0.0&&433.8&571.5&&624.6&867.0&&773.4&1097.2&&991.3&1434.6\\
b-s-corr\_06&500&0.0&c&1.00000&0.5&54.8&&248.7&154.4&&458.6&369.2&&\textbf{\underline{97.8}}&0.0&&101.6&3.9&&166.2&70.0\\
b-s-corr\_10&500&0.0&c&0.99998&0.4&26.1&&381.1&118.2&&2264.5&1196.4&&219.1&25.4&&228.3&30.7&&\textbf{\underline{174.7}}&0.0\\
uncorr\_01&1000&37.0&c&0.99999&0.4&66.6&&240.8&166.3&&\textbf{\underline{90.4}}&0.0&&115.0&27.2&&571.4&532.1&&168.9&86.9\\
uncorr\_06&1000&15.1&c&1.00000&0.5&39.1&&1293.3&4648.3&&\textbf{\underline{27.2}}&0.0&&37.3&37.1&&91.5&236.1&&46.6&71.1\\
uncorr\_10&1000&50.4&u&1.00000&0.2&11.7&&625.8&13008.4&&\textbf{\underline{4.8}}&0.0&&5.0&5.2&&10.9&127.3&&5.6&18.3\\
uncorr-s-w\_01&1000&19.7&c&0.99993&3355.4&88.4&&7158.2&97.8&&\textbf{\underline{3618.7}}&0.0&&3754.5&3.8&&-&-&&3946.9&9.1\\
uncorr-s-w\_06&1000&13.7&c&1.00000&0.5&45.1&&1187.7&9252.4&&\textbf{\underline{12.7}}&0.0&&13.3&4.7&&34.8&173.9&&17.0&33.9\\
uncorr-s-w\_10&1000&15.9&u&1.00000&0.2&16.5&&2162.6&28539.7&&\textbf{\underline{7.6}}&0.0&&7.9&4.3&&15.4&104.1&&8.5&13.0\\
b-s-corr\_01&1000&0.0&c&0.99996&1.4&93.0&&\textbf{\underline{456.5}}&0.0&&7802.0&1609.3&&6130.9&1243.1&&7992.7&1651.0&&9258.4&1928.3\\
b-s-corr\_06&1000&0.0&c&1.00000&0.3&55.2&&661.4&32.1&&38211.7&7531.9&&919.5&83.7&&1007.2&101.2&&\textbf{\underline{500.7}}&0.0\\
b-s-corr\_10&1000&0.0&c&0.99999&0.2&26.8&&874.5&30.0&&-&-&&1646.7&144.9&&1760.5&161.8&&\textbf{\underline{672.5}}&0.0\\
\hline
\end{tabular}
\end{adjustwidth}
}
\end{table}

\begin{table}[!htbp]
\centering
\caption{Details on the CP Search Performed by {\PWTBIB} on Small Size Instances}
\label{tab:Ares1}
{\tiny
\begin{adjustwidth}{-2.5cm}{}
\begin{tabular}{
@{\,}r@{\,}||
@{\,}r@{\,}|
@{\,}c@{\,}||
@{\,}r@{\,}|@{\,}r@{\,}|@{\,}r@{\,}
@{}r@{}||
@{\,}r@{\,}|@{\,}r@{\,}|@{\,}r@{\,}
@{}r@{}||
@{\,}r@{\,}|@{\,}r@{\,}|@{\,}r@{\,}
@{}r@{}||
@{\,}r@{\,}|@{\,}r@{\,}|@{\,}r@{\,}
}
\hline
\multirow{2}{*}{instance}&\multirow{2}{*}{m}&\multirow{2}{*}{{\textit{ver}}}&\multicolumn{3}{c}{\PWTBIBone}&&\multicolumn{3}{c}{\PWTBIBtwo}&&\multicolumn{3}{c}{\PWTBIBthree}&&\multicolumn{3}{c}{\PWTBIBfour}\\
\hhline{~~~---------------}
&&& 
$b$ & $f$ &\textit{{\ANKPu} runs}&&
$b$ & $f$ &\textit{{\ANKPu} runs}&& 
$b$ & $f$ &\textit{{\ANKPu} runs}&& 
$b$ & $f$ &\textit{{\ANKPu} runs} \\
\hline
\hline
\multicolumn{18}{c}{instance family \texttt{eil51}} \\
\hline
uncorr\_01&50&c&42&20&16$\mid$64 (25\%)&&42&20&16$\mid$64 (25\%)&&91&44&26$\mid$119 (22\%)&&42&20&16$\mid$64 (25\%)\\
uncorr\_06&50&c&34&17&17$\mid$65 (26\%)&&34&17&17$\mid$65 (26\%)&&72&35&31$\mid$101 (31\%)&&34&17&17$\mid$65 (26\%)\\
uncorr\_10&50&u&22&11&11$\mid$36 (31\%)&&22&11&11$\mid$36 (31\%)&&50&22&20$\mid$58 (35\%)&&22&11&11$\mid$36 (31\%)\\
uncorr-s-w\_01&50&c&46&21&14$\mid$84 (17\%)&&46&21&14$\mid$84 (17\%)&&163&79&41$\mid$199 (20\%)&&46&21&14$\mid$84 (17\%)\\
uncorr-s-w\_06&50&c&36&16&16$\mid$64 (25\%)&&36&16&16$\mid$64 (25\%)&&62&30&30$\mid$94 (32\%)&&36&16&16$\mid$64 (25\%)\\
uncorr-s-w\_10&50&u&8&4&4$\mid$33 (12\%)&&8&4&4$\mid$33 (12\%)&&26&12&12$\mid$54 (22\%)&&8&4&4$\mid$33 (12\%)\\
b-s-corr\_01&50&c&676&334&220$\mid$609 (36\%)&&676&334&220$\mid$609 (36\%)&&722&355&278$\mid$754 (37\%)&&676&334&220$\mid$609 (36\%)\\
b-s-corr\_06&50&c&100&50&50$\mid$139 (36\%)&&100&50&50$\mid$139 (36\%)&&104&51&61$\mid$169 (36\%)&&100&50&50$\mid$139 (36\%)\\
b-s-corr\_10&50&c&90&45&45$\mid$129 (35\%)&&90&45&45$\mid$129 (35\%)&&92&45&54$\mid$155 (35\%)&&90&45&45$\mid$129 (35\%)\\
uncorr\_01&250&c&110&52&50$\mid$254 (20\%)&&110&52&50$\mid$254 (20\%)&&293&138&97$\mid$472 (21\%)&&110&52&50$\mid$254 (20\%)\\
uncorr\_06&250&c&152&75&75$\mid$346 (22\%)&&130&64&64$\mid$324 (20\%)&&228&112&112$\mid$456 (24\%)&&130&64&64$\mid$324 (20\%)\\
uncorr\_10&250&u&36&18&18$\mid$145 (12\%)&&36&18&18$\mid$145 (12\%)&&151&74&73$\mid$264 (28\%)&&36&18&18$\mid$145 (12\%)\\
uncorr-s-w\_01&250&c&58&29&24$\mid$268 (9\%)&&58&29&24$\mid$268 (9\%)&&415&203&154$\mid$749 (21\%)&&58&29&24$\mid$268 (9\%)\\
uncorr-s-w\_06&250&c&48&23&23$\mid$255 (9\%)&&48&23&23$\mid$255 (9\%)&&175&86&85$\mid$415 (21\%)&&48&23&23$\mid$255 (9\%)\\
uncorr-s-w\_10&250&u&30&14&14$\mid$225 (6\%)&&30&14&14$\mid$225 (6\%)&&187&90&86$\mid$404 (21\%)&&30&14&14$\mid$225 (6\%)\\
b-s-corr\_01&250&c&600&296&265$\mid$1188 (22\%)&&600&296&265$\mid$1188 (22\%)&&684&338&366$\mid$1112 (33\%)&&600&296&265$\mid$1188 (22\%)\\
b-s-corr\_06&250&c&1148&572&567$\mid$1388 (41\%)&&662&331&331$\mid$885 (37\%)&&700&350&420$\mid$1104 (38\%)&&444&222&222$\mid$670 (33\%)\\
b-s-corr\_10&250&c&412&206&205$\mid$628 (33\%)&&232&116&116$\mid$452 (26\%)&&238&119&143$\mid$547 (26\%)&&232&116&116$\mid$452 (26\%)\\
uncorr\_01&500&c&1032&508&376$\mid$1080 (35\%)&&964&470&354$\mid$1034 (34\%)&&2292&1132&809$\mid$2482 (33\%)&&932&455&344$\mid$1014 (34\%)\\
uncorr\_06&500&c&332&164&163$\mid$755 (22\%)&&266&131&131$\mid$683 (19\%)&&610&301&300$\mid$1104 (27\%)&&266&131&131$\mid$683 (19\%)\\
uncorr\_10&500&u&36&18&18$\mid$285 (6\%)&&36&18&18$\mid$285 (6\%)&&269&131&128$\mid$548 (23\%)&&36&18&18$\mid$285 (6\%)\\
uncorr-s-w\_01&500&c&66&33&28$\mid$474 (6\%)&&66&33&28$\mid$474 (6\%)&&466&224&167$\mid$1160 (14\%)&&66&33&28$\mid$474 (6\%)\\
uncorr-s-w\_06&500&c&96&46&46$\mid$511 (9\%)&&94&45&45$\mid$509 (9\%)&&386&191&191$\mid$911 (21\%)&&94&45&45$\mid$509 (9\%)\\
uncorr-s-w\_10&500&u&38&17&16$\mid$442 (4\%)&&36&16&16$\mid$440 (4\%)&&382&187&187$\mid$872 (21\%)&&36&18&18$\mid$440 (4\%)\\
b-s-corr\_01&500&c&8318&4154&4061$\mid$9992 (41\%)&&4486&2238&2148$\mid$6128 (35\%)&&5310&2639&3034$\mid$7044 (43\%)&&4486&2238&2148$\mid$6128 (35\%)\\
b-s-corr\_06&500&c&21858&10918&10771$\mid$22543 (48\%)&&2804&1396&1391$\mid$3305 (42\%)&&3212&1601&1915$\mid$4433 (43\%)&&1960&976&976$\mid$2453 (40\%)\\
b-s-corr\_10&500&c&6402&3199&3167$\mid$6844 (46\%)&&628&314&313$\mid$1102 (28\%)&&650&324&388$\mid$1346 (29\%)&&354&177&177$\mid$830 (21\%)\\
\hline
\hline
\multicolumn{18}{c}{instance family \texttt{eil76}} \\
\hline
uncorr\_01&75&c&96&46&44$\mid$138 (32\%)&&96&46&44$\mid$138 (32\%)&&199&97&65$\mid$254 (25\%)&&96&46&44$\mid$138 (32\%)\\
uncorr\_06&75&c&56&26&26$\mid$107 (24\%)&&56&26&26$\mid$107 (24\%)&&91&44&44$\mid$149 (30\%)&&56&26&26$\mid$107 (24\%)\\
uncorr\_10&75&u&10&5&5$\mid$45 (11\%)&&10&5&5$\mid$45 (11\%)&&58&28&28$\mid$91 (30\%)&&10&5&5$\mid$45 (11\%)\\
uncorr-s-w\_01&75&c&110&54&33$\mid$170 (19\%)&&110&54&33$\mid$170 (19\%)&&338&167&86$\mid$433 (20\%)&&110&54&33$\mid$170 (19\%)\\
uncorr-s-w\_06&75&c&38&19&17$\mid$96 (18\%)&&38&19&17$\mid$96 (18\%)&&94&46&42$\mid$151 (28\%)&&38&19&17$\mid$96 (18\%)\\
uncorr-s-w\_10&75&u&20&10&10$\mid$74 (14\%)&&20&10&10$\mid$74 (14\%)&&65&29&28$\mid$121 (23\%)&&20&10&10$\mid$74 (14\%)\\
b-s-corr\_01&75&c&6218&3100&2684$\mid$6446 (42\%)&&6218&3100&2684$\mid$6446 (42\%)&&6856&3419&3592$\mid$8249 (44\%)&&6210&3096&2680$\mid$6437 (42\%)\\
b-s-corr\_06&75&c&110&55&55$\mid$178 (31\%)&&110&55&55$\mid$178 (31\%)&&128&63&76$\mid$233 (32\%)&&110&55&55$\mid$178 (31\%)\\
b-s-corr\_10&75&c&318&159&157$\mid$358 (44\%)&&224&112&112$\mid$272 (41\%)&&246&123&148$\mid$353 (42\%)&&174&87&87$\mid$226 (38\%)\\
uncorr\_01&375&c&212&106&102$\mid$438 (23\%)&&212&106&102$\mid$438 (23\%)&&473&236&178$\mid$785 (23\%)&&212&106&102$\mid$438 (23\%)\\
uncorr\_06&375&c&160&79&79$\mid$461 (17\%)&&160&79&79$\mid$461 (17\%)&&302&148&146$\mid$653 (22\%)&&160&79&79$\mid$461 (17\%)\\
uncorr\_10&375&u&46&22&21$\mid$230 (9\%)&&46&22&21$\mid$230 (9\%)&&545&268&260$\mid$718 (36\%)&&46&23&23$\mid$230 (10\%)\\
uncorr-s-w\_01&375&c&200&99&96$\mid$560 (17\%)&&186&92&89$\mid$545 (16\%)&&3624&1806&1386$\mid$4938 (28\%)&&186&92&89$\mid$545 (16\%)\\
uncorr-s-w\_06&375&c&78&38&37$\mid$437 (8\%)&&78&38&37$\mid$437 (8\%)&&266&127&126$\mid$661 (19\%)&&78&38&37$\mid$437 (8\%)\\
uncorr-s-w\_10&375&u&32&14&14$\mid$347 (4\%)&&30&15&15$\mid$345 (4\%)&&156&77&76$\mid$528 (14\%)&&30&15&15$\mid$345 (4\%)\\
b-s-corr\_01&375&c&568&278&227$\mid$1901 (12\%)&&568&278&227$\mid$1901 (12\%)&&662&325&318$\mid$1252 (25\%)&&568&278&227$\mid$1901 (12\%)\\
b-s-corr\_06&375&c&3360&1679&1667$\mid$3708 (45\%)&&414&207&207$\mid$771 (27\%)&&454&227&272$\mid$973 (28\%)&&414&207&207$\mid$771 (27\%)\\
b-s-corr\_10&375&c&5232&2610&2572$\mid$5543 (46\%)&&586&290&289$\mid$935 (31\%)&&668&331&396$\mid$1217 (33\%)&&472&233&233$\mid$822 (28\%)\\
uncorr\_01&750&c&488&243&238$\mid$991 (24\%)&&352&175&170$\mid$851 (20\%)&&1094&545&419$\mid$1924 (22\%)&&352&175&170$\mid$851 (20\%)\\
uncorr\_06&750&c&1254&624&610$\mid$1891 (32\%)&&348&172&172$\mid$970 (18\%)&&869&430&427$\mid$1610 (27\%)&&348&172&172$\mid$970 (18\%)\\
uncorr\_10&750&u&174&84&71$\mid$604 (12\%)&&92&46&46$\mid$510 (9\%)&&626&307&301$\mid$1091 (28\%)&&58&29&29$\mid$475 (6\%)\\
uncorr-s-w\_01&750&c&152&75&71$\mid$838 (8\%)&&150&74&70$\mid$836 (8\%)&&1668&824&630$\mid$3122 (20\%)&&150&74&70$\mid$836 (8\%)\\
uncorr-s-w\_06&750&c&136&65&62$\mid$803 (8\%)&&74&36&35$\mid$736 (5\%)&&384&191&187$\mid$1200 (16\%)&&74&36&35$\mid$736 (5\%)\\
uncorr-s-w\_10&750&u&28&13&13$\mid$677 (2\%)&&22&11&11$\mid$667 (2\%)&&550&274&271$\mid$1342 (20\%)&&22&11&11$\mid$667 (2\%)\\
b-s-corr\_01&750&c&2576&1282&1152$\mid$6124 (19\%)&&1754&871&744$\mid$5281 (14\%)&&2324&1150&1170$\mid$3725 (31\%)&&1754&871&744$\mid$5281 (14\%)\\
b-s-corr\_06&750&c&39336&19666&19490$\mid$39884 (49\%)&&1866&932&930$\mid$2597 (36\%)&&2158&1077&1290$\mid$3470 (37\%)&&686&343&343$\mid$1416 (24\%)\\
b-s-corr\_10&750&c&1664464&832213&819235$\mid$1652903 (50\%)&&4906&2450&2435$\mid$5635 (43\%)&&5288&2640&3150$\mid$7241 (44\%)&&1890&943&940$\mid$2615 (36\%)\\
\hline
\hline
\multicolumn{18}{c}{instance family \texttt{eil101}} \\
\hline
uncorr\_01&100&c&48&24&24$\mid$97 (25\%)&&48&24&24$\mid$97 (25\%)&&91&46&36$\mid$151 (24\%)&&48&24&24$\mid$97 (25\%)\\
uncorr\_06&100&c&90&42&42$\mid$161 (26\%)&&90&42&42$\mid$161 (26\%)&&151&72&72$\mid$228 (32\%)&&90&42&42$\mid$161 (26\%)\\
uncorr\_10&100&u&26&13&13$\mid$62 (21\%)&&26&13&13$\mid$62 (21\%)&&96&47&46$\mid$127 (36\%)&&24&12&12$\mid$60 (20\%)\\
uncorr-s-w\_01&100&c&44&21&12$\mid$121 (10\%)&&44&21&12$\mid$121 (10\%)&&149&73&34$\mid$269 (13\%)&&44&21&12$\mid$121 (10\%)\\
uncorr-s-w\_06&100&c&20&9&9$\mid$97 (9\%)&&20&9&9$\mid$97 (9\%)&&74&35&32$\mid$161 (20\%)&&20&9&9$\mid$97 (9\%)\\
uncorr-s-w\_10&100&u&22&11&11$\mid$105 (10\%)&&22&11&11$\mid$105 (10\%)&&67&31&30$\mid$158 (19\%)&&22&11&11$\mid$105 (10\%)\\
b-s-corr\_01&100&c&720&360&296$\mid$870 (34\%)&&720&360&296$\mid$870 (34\%)&&756&377&376$\mid$1015 (37\%)&&720&360&296$\mid$870 (34\%)\\
b-s-corr\_06&100&c&144&72&72$\mid$237 (30\%)&&144&72&72$\mid$237 (30\%)&&152&76&91$\mid$294 (31\%)&&144&72&72$\mid$237 (30\%)\\
b-s-corr\_10&100&c&582&291&289$\mid$658 (44\%)&&460&230&230$\mid$544 (42\%)&&466&233&280$\mid$660 (42\%)&&426&213&213$\mid$510 (42\%)\\
uncorr\_01&500&c&200&100&98$\mid$494 (20\%)&&200&100&98$\mid$494 (20\%)&&485&242&199$\mid$872 (23\%)&&200&100&98$\mid$494 (20\%)\\
uncorr\_06&500&c&196&98&98$\mid$606 (16\%)&&196&98&98$\mid$606 (16\%)&&360&176&176$\mid$840 (21\%)&&196&98&98$\mid$606 (16\%)\\
uncorr\_10&500&u&48&22&20$\mid$282 (7\%)&&44&22&21$\mid$278 (8\%)&&242&119&107$\mid$486 (22\%)&&44&22&22$\mid$278 (8\%)\\
uncorr-s-w\_01&500&c&626&308&253$\mid$1098 (23\%)&&612&301&247$\mid$1084 (23\%)&&27178&13577&10687$\mid$35473 (30\%)&&612&301&247$\mid$1084 (23\%)\\
uncorr-s-w\_06&500&c&68&33&33$\mid$488 (7\%)&&68&33&33$\mid$488 (7\%)&&300&148&144$\mid$815 (18\%)&&68&33&33$\mid$488 (7\%)\\
uncorr-s-w\_10&500&u&28&14&14$\mid$445 (3\%)&&28&14&14$\mid$445 (3\%)&&202&101&100$\mid$704 (14\%)&&28&14&14$\mid$445 (3\%)\\
b-s-corr\_01&500&c&6140&3065&2870$\mid$9999 (29\%)&&3732&1862&1702$\mid$7038 (24\%)&&4416&2203&2432$\mid$5966 (41\%)&&3706&1849&1689$\mid$7012 (24\%)\\
b-s-corr\_06&500&c&11490&5742&5695$\mid$11887 (48\%)&&714&357&356$\mid$1194 (30\%)&&758&379&454$\mid$1487 (31\%)&&670&335&334$\mid$1150 (29\%)\\
b-s-corr\_10&500&c&119750&59866&59499$\mid$115983 (51\%)&&4822&2408&2403$\mid$5161 (47\%)&&5044&2519&3017$\mid$6460 (47\%)&&2224&1111&1111$\mid$2642 (42\%)\\
uncorr\_01&1000&c&2148&1073&1063$\mid$2854 (37\%)&&1082&540&537$\mid$1726 (31\%)&&4006&2002&1679$\mid$5659 (30\%)&&896&447&444$\mid$1538 (29\%)\\
uncorr\_06&1000&c&986&491&486$\mid$1851 (26\%)&&750&374&371$\mid$1607 (23\%)&&1954&970&966$\mid$3011 (32\%)&&508&253&253$\mid$1357 (19\%)\\
uncorr\_10&1000&u&126&58&47$\mid$637 (7\%)&&76&38&37$\mid$572 (6\%)&&482&239&232$\mid$1057 (22\%)&&76&38&37$\mid$572 (6\%)\\
uncorr-s-w\_01&1000&c&32440&16204&12353$\mid$31686 (39\%)&&31940&15954&12152$\mid$31158 (39\%)&&-&-&-&&31826&15897&12107$\mid$31031 (39\%)\\
uncorr-s-w\_06&1000&c&376&185&172$\mid$1261 (14\%)&&96&47&47$\mid$955 (5\%)&&463&226&223$\mid$1530 (15\%)&&96&47&47$\mid$955 (5\%)\\
uncorr-s-w\_10&1000&u&36&17&15$\mid$875 (2\%)&&34&17&17$\mid$872 (2\%)&&350&175&174$\mid$1375 (13\%)&&34&17&17$\mid$872 (2\%)\\
b-s-corr\_01&1000&c&242728&121358&120894$\mid$252979 (48\%)&&73622&36807&36423$\mid$82106 (44\%)&&84676&42324&50052$\mid$103974 (48\%)&&65682&32837&32463$\mid$73997 (44\%)\\
b-s-corr\_06&1000&c&2087110&1043547&1025660$\mid$2115122 (48\%)&&7142&3570&3561$\mid$8126 (44\%)&&7758&3876&4640$\mid$10501 (44\%)&&1882&941&938$\mid$2867 (33\%)\\
b-s-corr\_10&1000&c&-&-&-&&35006&17496&17387$\mid$35797 (49\%)&&37988&18985&22651$\mid$46530 (49\%)&&7190&3589&3579$\mid$8131 (44\%)\\
\hline
\end{tabular}
\end{adjustwidth}
}
\end{table}

\subsection{Computational Experiments on the Set of Large Size Instances} \label{sec:ls}

The goal of our second experiment is to understand how fast {\ANKP} finds the approximate solutions of larger size instances and how efficient the pre-processing scheme is in this case. Solving large size instances turns out to be costly and requires computational capacity beyond that of an ordinary computer. Therefore, this series of experiments has been carried out on a computational cluster with 128 Gb RAM and 2.8 GHz 48-cores AMD Opteron processor. We use the same settings for the MIP solver as in our first experiment and set $\lambda=100$ for {\ANKP}. We investigate two families of the largest size instances of benchmark suite $B$, namely those based on the TSP problems \texttt{pla33810} and \texttt{pla85900} with $33810$ and $85900$ cities, respectively. Table \ref{tab:res2} reports the results. Columns 1-4 specify the instance's name, the total number of items $m$, the percentage of items discarded within the pre-processing step $\alpha$, and the version of the problem solved. The rest of the table describes the results for {\ANKPone} and for its two special cases: {\ANKPtwo} when {\ANKPone} is run without any pre-processing at all and {\ANKPthree} when {\ANKPone} does not use the reasoning based on the sequencing constraints to accelerate deduction of compulsory and unprofitable items when a problem in hand is unconstrained (see Section \ref{sec:RS} for details). Columns 5 and 10 provide the running time $t$ of {\ANKPone} (including the time taken by pre-processing) and of {\ANKPtwo}, respectively, while columns 6 and 11 give details on $\beta$, which is the percentage of auxiliary $y$-type variables that they use. The way to calculate $\alpha$ and $\beta$ is given in Section~\ref{sec:ss}. Columns 8 and 12 report $\omega$ as a relative gap in percents that compares the running times of {\ANKPone} and {\ANKPtwo}. Specifically, $\omega$ is computed as $\omega=100 \cdot \left(t^{A\!L\!G}-t^{M\!I\!N}\right)/t^{M\!I\!N}$, where $t^{A\!L\!G}$ is the running time of {\ANKPone} or {\ANKPtwo} and $t^{M\!I\!N}$ is the minimum of their running times. Column $t_{p}$ gives the pre-processing time taken by {\ANKPone}. Column $\gamma$ shows a ratio between the number of $y$-type variables used by {\ANKPtwo} and {\ANKPone}. Column $\rho$ gives a ratio between the upper bound obtained by {\ANKPu} with the same $\lambda=100$ and the lower bound {\ANKPone}. We do not provide runtime and some other results for {\ANKPthree} as it always performs worse than {\ANKPone}. For it, we only show the ratio $\eta$ between the pre-processing time of {\ANKPthree} and that one of {\ANKPone} to evaluate a speedup gained by utilizing the sequencing constraints for deduction of compulsory and unprofitable items in the case of {\NKPu}. Note that in {\ANKPthree} the pre-processing scheme decides whether an item is compulsory or unprofitable or none of these cases independently of other items whose properties are already known.

The results show that the pre-processing step is important and leads to great speeding-up. The pre-processing scheme excludes on average 27.7\% of items per instance of the {\U} type and 13.2\% of items per instance of the {\USW} type. The instances of these two types are vulnerable for reduction when the problem is {\NKPu} and when capacities are large. Furthermore, they are processed relatively fast as require much less number of auxiliary variables. Pre-processing allows {\ANKPone} to accelerate computations for the {\U} and {\USW} types of the instances by 328\% and 421\% on average, respectively. The average value of ratio $\gamma$ is 1.5 for the {\U} type against 1.2 for the {\USW} type. Interestingly, despite on the larger portion of auxiliary $y$-type variables excluded for the {\U} type, its speeding-up indicator $\omega$ is less than that one of the {\USW} type. Pre-processing is rather costly when applied to the unconstrained instances. However, the reasoning based on the sequencing constraints significantly improves the situation. This is shown by the values of $\eta$, which indicate that the time taken by pre-processing can be reduced up to $\sim$400 times when comparing {\ANKPone} to {\ANKPthree}. Otherwise, the running time of {\ANKPthree} often dominates one of {\ANKPtwo} that leads to avoiding the pre-processing stage when dealing with {\NKPu}.

In general, {\ANKPone} proves its ability to master large instances in a reasonable time. It needs less than $\sim$30 minutes to find an approximate solution to any instance of family \texttt{pla33810}. Almost all the instances of family \texttt{pla85900} can be solved approximately within 1 hour; it takes no longer than $\sim$4 hours for any of them. The quality of approximate solutions is outstanding as is confirmed by the very small values of ratio $\rho$.

\begin{table}[!htbp]
\centering
\caption{Results of Computational Experiments on Large Size Instances}
\label{tab:res2}
{\tiny
\begin{tabular}{
@{\,}r@{\,}||
@{\,}r@{\,}|
@{\,}r@{\,}|
@{\,}c@{\,}||
@{\,}r@{\,}|@{\,}r@{\,}|@{\,}r@{\,}|@{\,}r@{\,}|@{\,}r@{\,}
@{}r@{}||
@{\,}r@{\,}|@{\,}r@{\,}|@{\,}r@{\,}|@{\,}r@{\,}
@{}r@{}||
@{\,}r@{\,}
}
\hline
\multirow{2}{*}{instance} & \multirow{2}{*}{m} & \multirow{2}{*}{$\alpha$, \%} & \multirow{2}{*}{\textit{ver}} & \multicolumn{5}{c}{\ANKPone}&& \multicolumn{4}{c}{\ANKPtwo} && {\ANKPthree}\\
\hhline{~~~~------------}
&&&& 
$t$, sec & $\beta$, \% & $t_{p}$& $\omega$, \% & $\rho$ &&
$t$, sec & $\beta$, \% & $\omega$, \% & $\gamma$ && 
$\eta$ \\
\hline
\hline
\multicolumn{16}{c}{instance family \texttt{pla33810}} \\
\hline
uncorr\_01&33809&29.0&c&515&78.7&0&0&1.001&&3358&93.6&551&1.19&&1\\
uncorr\_06&33809&12.8&c&342&42.8&0&0&1.001&&2083&56.3&509&1.32&&1\\
uncorr\_10&33809&35.9&u&52&15.0&13&0&1.001&&334&26.4&543&1.76&&10\\
uncorr-s-w\_01&33809&19.3&c&435&89.5&0&0&1.001&&1411&93.5&225&1.04&&1\\
uncorr-s-w\_06&33809&11.2&c&607&47.7&0&0&1.001&&2795&56.2&361&1.18&&1\\
uncorr-s-w\_10&33809&8.7&c&25&18.2&0&0&1.001&&251&26.3&902&1.44&&1\\
b-s-corr\_01&33809&0.0&c&447&93.6&0&0&1.001&&463&93.6&4&1.00&&1\\
b-s-corr\_06&33809&0.0&c&566&56.3&0&0&1.001&&610&56.3&8&1.00&&1\\
b-s-corr\_10&33809&0.0&c&563&26.5&0&0&1.001&&603&26.5&7&1.00&&1\\
uncorr\_01&169045&30.6&c&587&76.6&0&0&1.001&&3050&93.6&419&1.22&&1\\
uncorr\_06&169045&12.8&c&1204&42.7&0&0&1.001&&2299&56.2&91&1.32&&1\\
uncorr\_10&169045&35.8&u&157&14.7&94&36&1.001&&115&26.4&0&1.79&&8\\
uncorr-s-w\_01&169045&15.2&c&348&90.5&0&0&1.001&&1561&93.5&348&1.03&&1\\
uncorr-s-w\_06&169045&11.7&c&590&47.2&0&0&1.001&&2141&56.2&263&1.19&&1\\
uncorr-s-w\_10&169045&9.0&c&549&18.0&0&0&1.001&&6245&26.3&1037&1.46&&1\\
b-s-corr\_01&169045&0.0&c&1384&93.7&0&0&1.001&&1410&93.7&2&1.00&&1\\
b-s-corr\_06&169045&0.0&c&438&56.4&0&0&1.002&&442&56.4&1&1.00&&1\\
b-s-corr\_10&169045&0.0&c&718&26.3&0&0&1.002&&814&26.3&13&1.00&&1\\
uncorr\_01&338090&31.6&c&1589&75.4&0&0&1.001&&9315&93.5&486&1.24&&1\\
uncorr\_06&338090&12.8&c&1023&42.6&0&0&1.001&&1645&56.2&61&1.32&&1\\
uncorr\_10&338090&35.9&u&947&14.7&238&0&1.001&&1224&26.3&29&1.79&&6\\
uncorr-s-w\_01&338090&15.2&c&1209&90.5&0&0&1.001&&5125&93.5&324&1.03&&1\\
uncorr-s-w\_06&338090&11.9&c&966&47.1&0&0&1.001&&2033&56.2&111&1.19&&1\\
uncorr-s-w\_10&338090&9.0&c&1156&18.0&0&0&1.001&&3621&26.3&213&1.46&&1\\
b-s-corr\_01&338090&0.0&c&829&93.6&0&0&1.001&&857&93.6&3&1.00&&1\\
b-s-corr\_06&338090&0.0&c&873&56.3&0&0&1.002&&947&56.3&8&1.00&&1\\
b-s-corr\_10&338090&0.0&c&1095&26.3&0&0&1.002&&1176&26.3&7&1.00&&1\\
\hline 
\hline
\multicolumn{16}{c}{instance family \texttt{pla85900}} \\
\hline
uncorr\_01&85899&32.4&c&2514&73.8&0&0&1.002&&10614&93.5&322&1.27&&1\\
uncorr\_06&85899&13.5&c&3028&41.6&0&0&1.002&&40283&56.1&1230&1.35&&1\\
uncorr\_10&85899&40.8&u&213&13.7&59&0&1.002&&323&26.3&52&1.92&&28\\
uncorr-s-w\_01&85899&16.4&c&1683&88.6&0&0&1.002&&9566&93.5&468&1.06&&3\\
uncorr-s-w\_06&85899&12.3&c&1985&46.6&0&0&1.002&&13408&56.2&575&1.20&&1\\
uncorr-s-w\_10&85899&13.6&u&158&17.2&3&0&1.002&&1165&26.3&637&1.53&&393\\
b-s-corr\_01&85899&0.0&c&3967&93.6&0&2&1.002&&3903&93.6&0&1.00&&1\\
b-s-corr\_06&85899&0.0&c&1570&56.3&0&0&1.002&&1575&56.3&0&1.00&&1\\
b-s-corr\_10&85899&0.0&c&3491&26.4&0&0&1.002&&3627&26.4&4&1.00&&1\\
uncorr\_01&429495&32.5&c&3436&73.6&0&0&1.002&&27833&93.5&710&1.27&&1\\
uncorr\_06&429495&13.6&c&5495&41.4&0&0&1.002&&47092&56.2&757&1.36&&1\\
uncorr\_10&429495&40.4&u&1471&13.8&936&58&1.001&&934&26.3&0&1.90&&18\\
uncorr-s-w\_01&429495&16.3&c&2380&90.2&0&0&1.002&&20790&93.5&773&1.04&&1\\
uncorr-s-w\_06&429495&12.8&c&4508&46.3&0&0&1.002&&10825&56.2&140&1.21&&1\\
uncorr-s-w\_10&429495&13.2&u&545&17.4&23&0&1.001&&1311&26.3&141&1.51&&256\\
b-s-corr\_01&429495&0.0&c&3105&93.6&0&0&1.002&&3145&93.6&1&1.00&&1\\
b-s-corr\_06&429495&0.0&c&5483&56.2&0&0&1.002&&5457&56.2&0&1.00&&1\\
b-s-corr\_10&429495&0.0&c&4751&26.3&0&0&1.002&&4872&26.3&3&1.00&&1\\
uncorr\_01&858990&33.2&c&6904&72.6&0&0&1.002&&11982&93.5&74&1.29&&1\\
uncorr\_06&858990&13.6&c&5510&41.4&0&0&1.002&&7860&56.2&43&1.36&&1\\
uncorr\_10&858990&40.6&u&3650&13.9&1544&0&1.001&&4627&26.3&27&1.90&&16\\
uncorr-s-w\_01&858990&16.4&c&4859&90.2&0&0&1.002&&43064&92.9&786&1.03&&1\\
uncorr-s-w\_06&858990&12.7&c&7095&46.3&0&0&1.002&&15904&56.2&124&1.21&&1\\
uncorr-s-w\_10&858990&13.2&u&5058&17.4&68&0&1.001&&12868&26.3&154&1.51&&183\\
b-s-corr\_01&858990&0.0&c&5474&93.5&0&0&1.002&&5498&93.5&0&1.00&&1\\
b-s-corr\_06&858990&0.0&c&12566&56.2&0&0&1.002&&12681&56.2&1&1.00&&1\\
b-s-corr\_10&858990&0.0&c&13806&26.4&0&0&1.002&&13981&26.4&1&1.00&&1\\
\hline
\hline
\end{tabular}
}
\end{table}

\section{Conclusion}

We have introduced a new non-linear knapsack problem where items to be selected are subject to the total reward that a vehicle obtains by summing up the profits of chosen items and the subtracting costs resulted from their transportation along a fixed route. We have shown that both the constrained and unconstrained versions of the problem are $\mathcal{NP}$-hard. Our proposed pre-processing scheme can significantly decrease the size of instances making them easier for computation. The experimental results show that small size instances can be solved to optimality in a reasonable time by any of the two proposed exact approaches. Larger instances can be efficiently handled by our approximate approach producing near-optimal solutions. 

As a future work, this problem has several natural generalizations. The first evident generalization is for sure the traveling thief problem where the sequence of cities may be changed. This variant asks for the mutual solution of the traveling salesman and knapsack problems. Another interesting situation takes place when cities may be skipped because are of no worth, for example any item stored there has in fact low or negative contribution to the total reward. Finally, the possibility to pickup and deliver the items is for certain one another challenging problem. The outcomes of our research can further be adopted to solve routing problems with nonlinear cost functions, for example those when such a measure as gallon per vehicle mile versus load is used.

\section{Acknowledgements} 
We want to thank the referees for their valuable suggestions which helped to improve the paper.
This research has been supported through ARC Discovery Project DP130104395.

\section{References}

\bibliographystyle{model5-names}
\bibliography{references}

\end{document}